\documentclass[acmsmall]{acmart}
\usepackage{pifont}
\usepackage{tikz}
\usepackage{wasysym}
 \usepackage{multirow}
\usepackage{booktabs}
\usepackage{utfsym}
\usepackage{array}
\usepackage{fontawesome}
\usepackage{threeparttable}
\usepackage{makecell}

\AtBeginDocument{%
  }

\setcopyright{acmlicensed}
\copyrightyear{2018}
\acmYear{2018}
\acmDOI{XXXXXXX.XXXXXXX}

\acmJournal{JACM}
\acmVolume{37}
\acmNumber{4}
\acmArticle{111}
\acmMonth{8}

\begin{document}

\title{Image Privacy Protection: A Survey}

\author{Wenying Wen}

\email{wenyingwen@sina.cn}
\author{Ziye Yuan}
\author{Yushu Zhang}
\author{Tao Wang}
\author{Xiangli Xiao}
\author{Ruoyu Zhao}
\author{Yuming Fang}

\affiliation{%
  \institution{Jiangxi University of Finance and Economics}
  \city{Nanchang}
  \state{Ohio}
  \country{China}
}

\authorsaddresses{Authors' addresses: Wenying Wen, Ziye Yuan, Yushu Zhang, Tao Wang, Xiangli Xiao, Ruoyu Zhao, and Yuming Fang are with the School of Computing and Artificial Intelligence, Jiangxi University of Finance and Economics, Nanchang, China (e-mail: wenyingwen@sina.cn, yzy$\_$stu@sina.com, yushu@nuaa.edu.cn, wangtao21@nuaa.edu.cn, xiaoxiangli@nuaa.edu.cn, zhaoruoyu@nuaa.edu.cn, and leo.fangyuming@foxmail.com).}

\renewcommand{\shortauthors}{Wen et al.}

\begin{abstract}
Images serve as a crucial medium for communication, presenting information in a visually engaging format that facilitates rapid comprehension of key points. Meanwhile, it is important to note that during the processes of transmission and storage, images also contain significant amounts of sensitive information. If not managed properly, this information may be vulnerable to exploitation for personal gain, potentially infringing on privacy rights and other legal entitlements. Consequently, researchers continue to propose some approaches for preserving image privacy and publish reviews that provide comprehensive and methodical summaries of these approaches. However, existing reviews tend to categorize either by specific scenarios, e.g., image privacy concerns in social networks and intelligent surveillance systems for video monitoring and environmental assistive living, or by specific privacy objectives, such as face. This classification somewhat restricts the reader's ability to grasp a holistic view of image privacy protection and poses challenges in developing a total understanding of the subject that transcends different scenarios and privacy objectives. Instead of examining image privacy protection from a single aspect, it is more desirable to consider user needs for a comprehensive understanding. To fill this gap, we conduct a systematic review of image privacy protection approaches based on privacy protection goals. Specifically, we define the attribute known as privacy-sensitive domains and use it as the core classification dimension to construct a comprehensive framework for image privacy protection that encompasses various scenarios and privacy objectives. This framework offers a deep understanding of the multi-layered aspects of image privacy, categorizing its protection into three primary levels: data-level image privacy protection, content-level image privacy protection, and feature-level image privacy protection. For each category, we analyze the main approaches and features of image privacy protection and systematically review representative solutions. Finally, we discuss the challenges and future directions of image privacy protection.

\end{abstract}




\maketitle

\section{INTRODUCTION}
\subsection{Background}
The image is a term that has evolved over time in the evolution of history and the development of language. The ancient Greek philosopher Aristotle, in his book Metaphysics, emphasized the importance of experience in the acquisition of knowledge \cite{139}, which is often acquired through the senses (including vision). In this process, images, as products of visual perception, undoubtedly play an important role in the formation and accumulation of knowledge. From ancient wall paintings to modern multimedia digital images, images have always conveyed a huge amount of information to us by being intuitive, rich, and varied. Nowadays, images are one of the most important medias for people to express their views, opinions, and emotions to each other \cite{140}. After the development of human society to a certain stage, although the text has also become an important means of transmitting information \cite{141}, but the expressive power of the text is still relatively weak compared to images \cite{142}. For example, as John Berg said \cite{143}, “Viewing precedes speech. We are viewers before anything else”.  His words point to the priority of images in perceiving and understanding the world. The power of images to convey information and express ideas is widely recognized and accepted, confirming the concept that “a picture is worth a thousand words” \cite{144}. For instance, 
In news reporting, a photo depicting the cruelty of war and the exhaustion of soldiers can visualize the pain of war and the hardship of soldiers more than a thousand words.

With the rapid advancement of smartphones and other mobile devices, various devices capable of collecting high-quality images have been successively invented and utilized. Consequently, the volume of image transmission is growing at a phenomenal rate and the range of applications is expanding significantly. For example, facebook, 35 billion photos are uploaded to it everyday. That's 250,000 times per minute or 4,000 times per second\footnote{https://www.websiterating.com/zh-CN/blog/research/facebook-statistics/references}. Since the launch of Instagram, more than 5 billion pictures and videos have been shared on the platform. It is reported that 1,074 images are uploaded to Instagram every second\footnote{https://www.websiterating.com/zh-CN/blog/research/instagram-statistics/references}. In addition, the telemedicine sharing platform proposes a fast and convenient diagnostic service to realize cross-regional cooperation among doctors. Most patients, especially those in areas where medical resources are scarce, are more inclined to share their medical images on this platform in order to be diagnosed and treated by specialists. Especially after the outbreak of New Crown Pneumonia in the early of 2020\footnote{https://www.peopleapp.com/column/30036878807-500001930699}, the telemedicine sharing platform turned out to be the main consultation tool as the rapid spread of the virus made it impossible for specialists from all over the world to gather for consultations. A large number of medical images of the patient's lungs, such as X-rays \cite{145}, are disseminated on the platform, making it convenient for medical specialists everywhere to scrutinize them. In addition to this, a large number of images are required for transmission and analysis in areas, e.g., intelligent surveillance, aerial reconnaissance, and light-field imaging.

While images demonstrate their unique value in various domains, they also pose potential safety hazards, i.e., the information they carry may reveal the privacy of individuals. In other words, when the sending end transmits an image, it mainly conveys specific information to the receiving end. At the same time, private information in the image content may be actively or passively exposed to others during transmission or storage. This includes biological privacy information, e.g., gender, race, age, and body condition, as well as non-biological privacy information, e.g., geographical location, buildings, and weather. For example, on October 13, 2023, three elite Tanzanian athletes sued MultiChoice (T) Ltd for invasion of privacy and for using their likenesses without their consent.MultiChoice (T) Ltd used their likenesses to entice members of the public to subscribe to its service to watch athletes competing in the Olympics. The images were allegedly posted on various channels, billboards across the country, and on MultiChoice (T) Ltd's own official Instagram and Twitter accounts\footnote{https://bowmanslaw.com/insights/tanzania-privacy-rights-and-image-exploitation-an-analysis-of-tanzanian-high-court-civil-case-no-15-of-2021-multichoice-t-ltd-vs-alphonce-felix-simbu-and-2-others/}. In addition, at the end of February 2023, a ransomware attack on the U.S. Marshals Service sparked speculation. According to the hackers, the stolen files included photographs and aerial photos of military bases and other heavily guarded areas, copies of passports and identification documents, as well as information about wiretapping and tracking personnel listening data, drone footage from military bases, and more\footnote{https://tech.news.am/eng/news/982}. Therefore, protecting the privacy of images is crucial, considering the absence of a delete button on the internet!

Image privacy includes not only common facial features, but also multiple dimensions such as personal information, geographic location, environment, and objects. Therefore it is a multidimensional and complex structure that derives from its multi-layered nature of information, different socio-cultural contexts, changing technological environments, and individual differences in privacy perceptions. Here, inspired by the review of Liu et al. \cite{2}, we categorize privacy into three categories: contextual privacy, observable privacy, and machine privacy.
\begin{itemize}

\item {\itshape{Contextual privacy:}} It refers to correlation analysis in complex contextual environments through data from other modalities, which may reveal specific content in the current modal data. Such cross-modal data correlation analysis may infer sensitive confidential information through indirect means, posing a threat to image privacy.

\item {\itshape{Observable privacy:}} It refers to the private content of an image that is visible to the naked eye, usually including specific targets. This type of privacy content is displayed directly in view, typically maintaining a consistent format, and is capable of conveying complete visual information to a trusted user. However, once it falls into the hands of an attacker, it can lead to the full disclosure of the information, creating a serious privacy risk. 

\item {\itshape{Machine privacy:}} It refers to those private information that can be inferred or detected by machines, which are often difficult for human observers to detect directly. They are usually hidden in certain features or attributes of an image and can only be revealed by utilizing machine detection and inference techniques in the latent feature space \cite{132,133}. 
\end{itemize}

\subsection{The Keys to Image Privacy Protection}
Devices capable of acquiring images are everywhere today. According to the Global Mobile Market Report 2021\footnote{Newzoo (2021), Global Mobile Market Report 2021}, there are more than 4 billion smartphone users worldwide, with China leading the way with a total of 954 million smartphone users and a penetration rate of 66$\%$. Images can be taken and published wherever, whenever and with just a few clicks. However, if such images are released freely and publicly without proper processing, they can easily leave those associated with the image in a passive state, as if they are being pried into some of their secrets. This privacy issue has been of widespread concern to governments across the globe\footnote{https://www.apc.org/en/pubs/protect-privacy-digital-age-world-governments-can-and-must-do-more}, and various privacy policies have been introduced in the hope of further reducing the risk of privacy breaches. For instance, the European Union (EU) adopted the General Data Protection Regulation (GDPR)\footnote{https://gdpr-info.eu} in 2016, which defines data protection and security rules across Europe and is expected to impose hefty fines for non-compliance. Similar legislative developments are also underway in the US, such as in California, where the California Consumer Privacy Act (CCPA)\footnote{https://leginfo.legislature.ca.gov/faces/codes$\_$displayText.xhtml?division=3.$\&$part=4.$\&$lawCode=CIV$\&$title=1.81.5} came into effect in 2018 and is considered by many to be a potential model for a US-wide data privacy law\footnote{https://www.thalesgroup.com/en/markets/digital-identity-and-security/government/biometrics/biometric-data}, and a variety of other privacy protection schemes have emerged. In addition to legal policy measures, a number of privacy protection schemes are also emerging in the academic world. However, regardless of the specific scenarios or advanced technologies, the key of image privacy protection is always to identify the level of privacy protection that the privacy subject expects and what they want to protect.

\textbf{Case study:	Whether image owners truly understand and are aware of the private content that may be contained in their images?}

In the context of online social networking, Alice wants to share a photo. This photo contains many things, such as faces or buildings. In order to prevent certain essential information from being compromised, the privacy party must surely first consider what content needs to be protected. While image full encryption gives her greater assurance by erasing the entire content compared to most other methods. But there is no doubt that it defeats the purpose and fun of sharing photos. Consequently, accurately defining and protecting sensitive areas within the image becomes paramount. However, the process is not simple, as a photo often contains a wealth of information. From color schemes and clothing choices to background settings and weather conditions, and even facial expressions, all of these may inadvertently reveal subtle details about a person's personal life. As such, before implementing privacy protection measures, Alice needs to deliberate carefully and clarify her privacy objectives.

Indeed, the above case is merely a simplified reflection of privacy protection issues, and the complexity of real-world scenarios far exceeds this, with privacy protection considerations being even broader and deeper. The understanding of privacy varies greatly among individuals, not only in terms of the definition of privacy content but also in the level of privacy protection expectations, as shown in Fig. 1. In view of this, in the following Sections 3 to 5, we will comprehensively and thoroughly explore all aspects of image privacy protection, striving to present readers with a more complete and in-depth understanding framework.

\subsection{Motivation}
The above analysis and case studies indicate that the handling of image privacy protection is vast and complex, and there is an urgent need for more advanced privacy protection technologies to give the public a sense of security. Therefore, the research community has explored more effective solutions to different problems separately to meet the needs of privacy protection. However, one issue remains: existing work primarily considers problems from an isolated perspective, addressing each privacy concern individually. There is an urgent need for a comprehensive analysis of these methods to provide fundamental insights into the field and facilitate the establishment of a privacy-friendly image sharing environment. To this end, we review the literature published on this topic in recent years and establish a comprehensive image privacy protection framework based on privacy objectives. The framework covers most of the different privacy goals and privacy protection issues considered, and offers a comparative analysis of typical cases in each category.

\begin{figure}[h]
  \centering
  \includegraphics[scale=0.35]{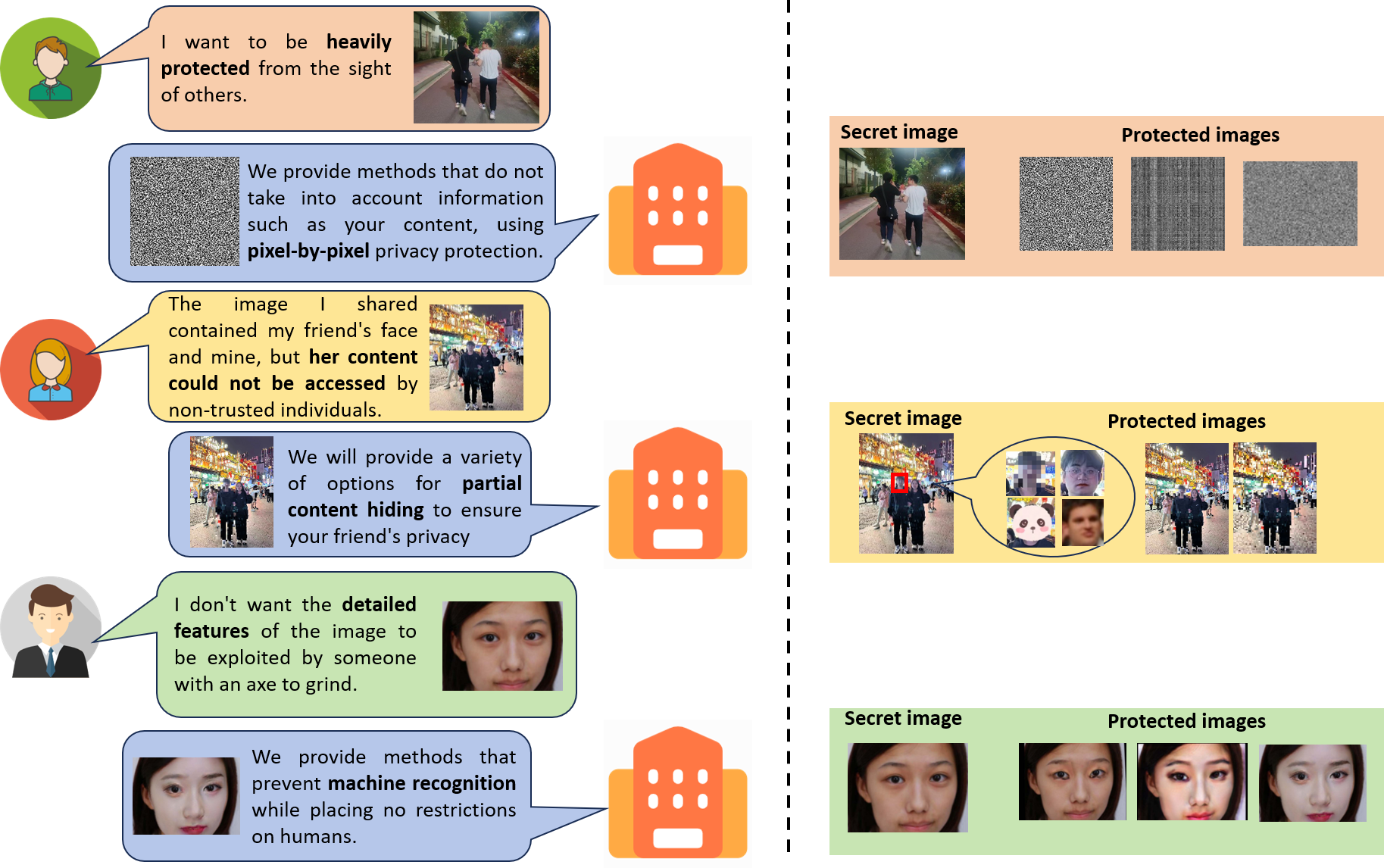}
  \caption{Differences in privacy requirements among different individuals.}
\end{figure}

\subsection{Analysis of Existing Surveys}	 
There are several reviews related to this topic, including the following contents:
\begin{itemize}

\item {\itshape{Specific scenarios:}} These literatures focus solely on privacy issues within specific contexts. Liu et al. \cite{2} investigated “privacy intelligence” from a user-centric perspective, addressing modern privacy concerns in dynamic online social network (OSN) image sharing. They present key attributes and classifications of OSN image privacy, along with an advanced privacy analysis framework based on the lifecycle of OSN image sharing. Alemany et al. \cite{3} provided an in-depth analysis of privacy decision-making mechanisms within OSNs. Padilla-López et al. \cite{4} surveyed existing privacy-aware intelligent surveillance systems and provided valuable discussions on critical aspects of visual privacy. Winkler et al. \cite{7} outlined the characteristics of visual sensor network (VSN) applications, including the security threats, attack scenarios, and major security challenges involved. They identified and discussed various security requirements, provided an extensive overview of related work within each category, discussed privacy protection techniques, and identified trends in VSN security and privacy.

\item {\itshape{Specific objectives:}} These issues are analyzed with designated privacy goals in mind. Wu et al. \cite{5} discussed the latest advancements in facial identity concealment technologies, emphasizing privacy protection methods that hide or safeguard facial biometric data prior to capture by camera devices. They evaluate the relative performance of facial privacy protection methods and identify open challenges and future work to consider in this research area. Wu et al. \cite{6} comprehensively introduced privacy-related research in the field of biometrics and reviewed existing works on biometric privacy enhancement techniques applied to facial biometrics. To improve scientific research with large datasets for patient care while preventing patient privacy breaches, Wu et al. \cite{8} outlined current and next-generation joint, secure, and privacy-preserving artificial intelligence approaches for medical imaging.
\end{itemize}

The previous reviews often focused on specific scenarios (intelligent surveillance or online social networks) or specific privacy objectives (biometrics or medical images). While these image privacy reviews indeed provide thorough analyses of privacy issues and protection methods within fixed scenarios or targeted objectives, image privacy, as a broader concept, encompasses scenarios and contents that extend beyond what these reviews cover. 


Two key points distinguish our review from previous works. First, in terms of the breadth of research content, this review is not limit to a particular scenario or privacy protection goal. Rather, it covers a wide range of privacy issues that may be encountered in various scenarios and the solutions proposed for different privacy contents. As a result, most of the challenges faced in the field of image privacy and their response strategies can be easily located in the framework of this review. Second, this review pioneers a new perspective privacy sensitive domains as a basis for categorizing image privacy protection schemes. In this framework, privacy-sensitive domains are carefully categorized into three main categories: pixels domain, content domain, and feature domain. This categorization approach focuses on categorizing privacy objects according to their categories rather than directly targeting the specific privacy objects themselves, such as faces or medical lesions. 

In summary, our review offers a comprehensive understanding of image privacy through the following contributions: 1) Property and classification of image privacy; 2) A novel framework for image privacy protection; 3) A positioning system within the framework for any image privacy protection scheme; 4) Potential solutions for privacy issues across any privacy sensitive domain; 5) A thorough review of current advancements in image privacy protection solutions; 6) Discussion of challenges and future directions for image privacy protection.

The remainder of this survey is organized as follows. Section 2 describes the different levels of privacy protection options. Sections 3 to 5 focus on each category, review representative solutions, and discuss common principles. Section 6 discusses the challenges and future directions of image privacy protection. Section 7 is a brief conclusion.

\section{IMAGE PRIVACY PROTECTION: AN OVERVIEW}
\subsection{Property and Taxonomy}
The core essence of image privacy protection lies in ensuring that sensitive information remains both securely concealed and naturally flowing, thereby constructing an impenetrable barrier against any unauthorized attempts at prying or acquisition. Given this, researchers closely focus on the actual sensitive content that users wish to protect and tailor privacy protection frameworks to precisely meet user needs. Building upon previous privacy protection cases, We define privacy-sensitive domains as key pillars of a privacy-preserving scheme.

\textbf{Privacy-sensitive domain.} The privacy sensitive domain refers to the privacy region in an image that needs to be protected, covering all parts of the image that involve sensitive information. Due to the different hierarchical nature of image information, the focus of different privacy protection schemes varies. The privacy-sensitive domain defined in this review effectively categorizes the privacy content and provides support for the framework design of subsequent privacy protection schemes. Specifically, the privacy-sensitive domains include the pixel domain, the visual content domain, and the feature domain. These three types of domains can adequately cover the privacy goals involved in existing privacy-preserving schemes. In addition, the protection schemes designed for these three types of privacy-sensitive domains present different levels of visual usability, i.e., no visual usability, general visual usability, and high visual usability, in order to meet the privacy protection needs in different scenarios.

Here, we classify image privacy protection based on the privacy-sensitive domain into three categories from low to high levels, as shown in Fig. 2.

\begin{figure}[h]
  \centering
  \includegraphics[scale=0.35]{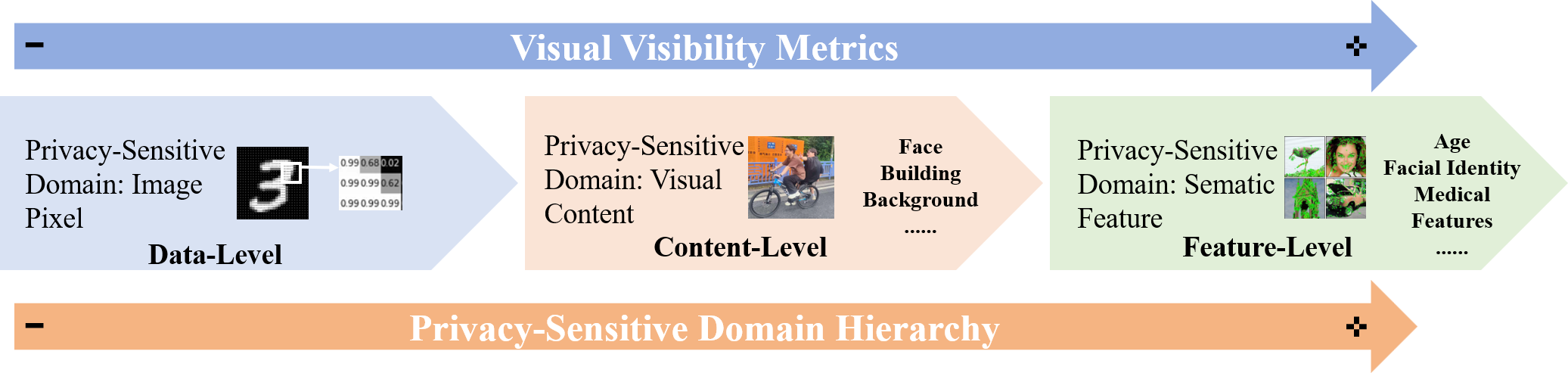}
  \caption{Different categories of image privacy protection.}
\end{figure}

\begin{itemize}
\item {\itshape{Data-level privacy protection:}} It focuses on elevating the entire image to an impeccable level of security by meticulously manipulating every pixel value, ensuring that no trace of information is leaked to unauthorized observers. The core of this protection mechanism lies in its comprehensiveness, as it does not rely on identifying and analyzing specific sensitive elements within the image content but treats the entire image as an indivisible privacy-sensitive entity. In other words, it abandons the traditional approach of protecting only part of the image content and adopts a more thorough and straightforward security measure. 

\item {\itshape{Content-level privacy protection:}} It primarily involves accurately defining privacy-sensitive areas in images, which may encompass specific parts of the image such as faces, buildings, license plate numbers, or even the entire image as sensitive. For these sensitive areas, targeted content modification techniques are employed, aiming to safeguard privacy information from unauthorized access while ensuring that the modified image retains sufficient visual content to facilitate rapid recognition of its core information by authorized users or individuals with relevant prior knowledge. The crux of this strategy lies in striking a balance between privacy protection and image usability, effectively guarding against potential information theft risks while preserving the image's fundamental function as a medium for information transmission. 

\item {\itshape{Feature-level privacy protection:}} It aims to implement effective protection measures for various detailed features contained in images. These features not only encompass sensitive biometric information, such as facial recognition features and personal identity details, but also include non-biometric data such as license plate numbers, which may not be immediately apparent to the human eye but are crucial elements easily captured and analyzed by machine learning algorithms and automatic recognition systems. Through feature-level privacy protection, it is possible to effectively prevent these sensitive features from being easily identified and utilized by specific machines or intelligent systems. Some solutions can not only effectively resist precise capture by machine recognition but also ingeniously prevent human snooping.

\end{itemize}

\subsection{Image Privacy Protection Framework}
We design a framework for image privacy protection based on privacy-sensitive domains to assist in identifying various privacy issues within images and researching corresponding countermeasures. This framework encompasses all privacy-sensitive domains in images, ensuring that any image privacy protection solution can find its appropriate classification within this framework, as illustrated in Fig. 3.

\begin{figure}[h]
  \centering
  \includegraphics[scale=0.3]{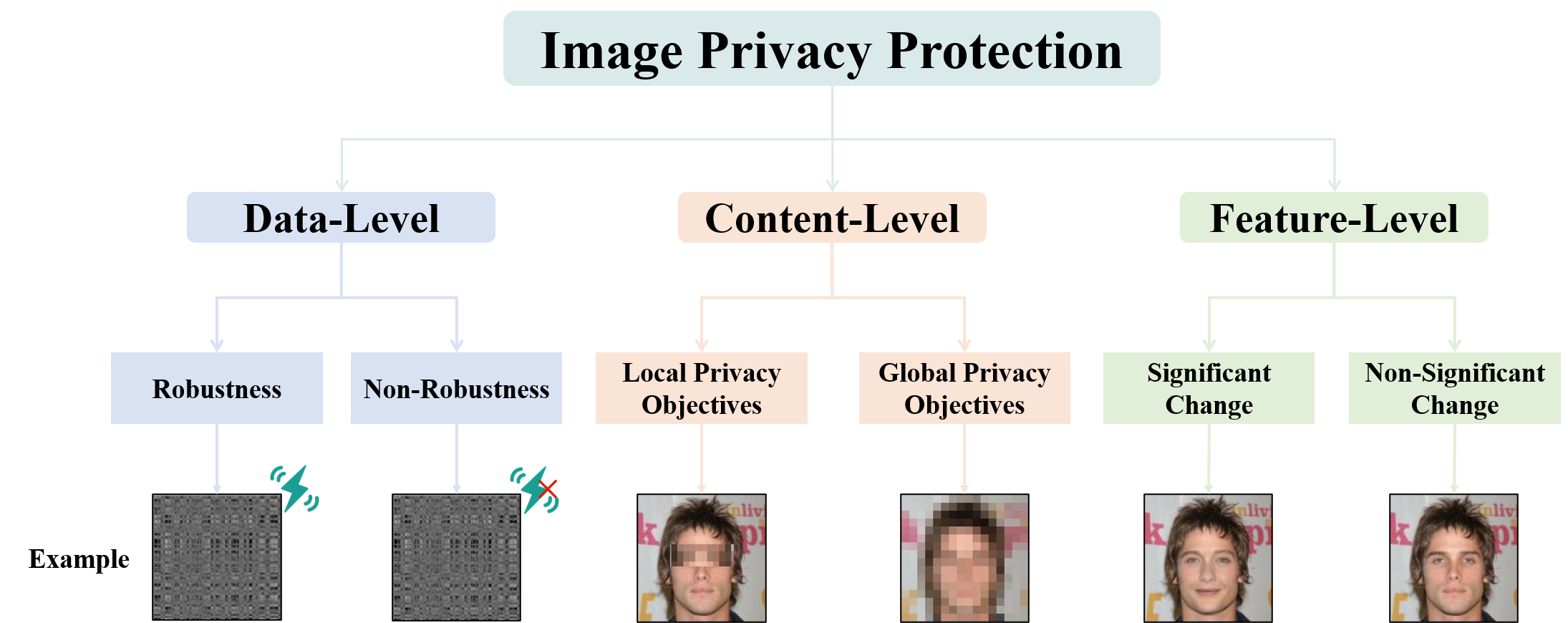}
  \caption{A framework of image privacy protection.}
\end{figure}

\subsubsection{Data-Level Privacy Protection}

\begin{itemize}
\item {\bfseries{Robustness.}} This kind of scheme aims to ensure the security and recoverability of image information, i.e., not only to ensure the concealment of image content, but also to have the ability to effectively reconstruct the secret image after suffering from malicious attacks or data damage, so as to ensure the availability of data.

\item {\bfseries{Non-robustness.}} The core of this kind of scheme lies in ensuring that even if the image data is attacked maliciously, it is impossible to easily or completely reconstruct the complete information of the original image. This is to create a tighter privacy protection barrier that makes any unauthorized attempt to recover the image content extremely difficult or impossible.

\end{itemize}

\subsubsection{Content-Level Privacy Protection}

\begin{itemize}
\item {\bfseries{Local privacy objectives.}}  This type of solution focuses on precisely modifying sensitive content in images while preserving other non-sensitive information and aesthetic elements, ensuring that the modified image remains visually appealing and attractive.

\item {\bfseries{Global privacy objectives.}}  This type of solution considers the overall content of the image as the core objective of privacy protection, while maintaining a certain composition, color scheme, and atmosphere. This ensures that the image still has the ability to transmit information, while privacy is protected.
\end{itemize}

\subsubsection{Feature-Level Privacy Protection}

\begin{itemize}
\item {\bfseries{Significant change.}} This type of approach focuses on significantly adjusting and transforming the key features within an image. Its primary objective is to fundamentally disrupt the possibility of artificial intelligence algorithms capturing and utilizing these features, while ensuring that these alterations pose sufficient obstacles to the intuitive recognition abilities of human observers. 

\item {\bfseries{Non-significant change.}} This type of approach specializes in detailed privacy protection processing for the intricate features within an image, while ensuring that this process does not disturb or diminish humans' overall understanding and perception of the original content of the image.

\end{itemize}

\section{DATA-LEVEL IMAGE PRIVACY PROTECTION}
This section provides a detailed analysis of data-level image privacy protection schemes, encompassing their abilitys, primary methodologies, and shared design principles. Fig. 4 offers an overview of the analysis conducted during this phase.

\begin{figure}[h]
  \centering
  \includegraphics[scale=0.4]{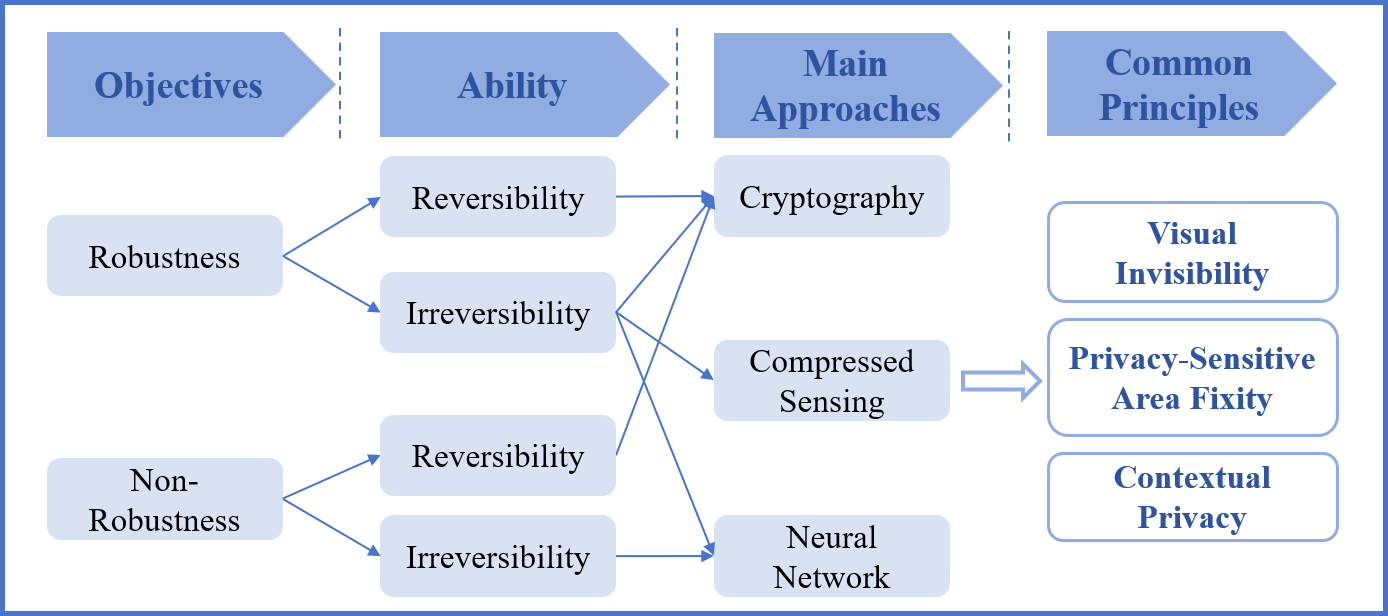}
  \caption{Overview of the data-level image privacy protection.}
\end{figure}

\subsection{Abilitys of Data-Level Privacy Protection}
Data-level image privacy protection strategies achieve an ultimate level of pixel-by-pixel security, ensuring that every data point is rigorously safeguarded. However, such intense privacy measures inevitably impact the usability of the image, rendering the processed image virtually uninformative in its encrypted state. The ideal encryption effect is that the image is completely unrecognizable during transmission and storage, and can only be losslessly restored to its full and clear content by the intended recipient through specific decryption means. Unfortunately, not all existing solutions perfectly achieve this goal, with some potentially compromising the integrity or clarity of the image post-decryption. Therefore, regardless of whether a data-level privacy protection scheme possesses robustness, we should consider the following two abilitys separately.

(1) Data-Level Reversibility. In the context of specific sensitive fields, particularly medical image processing, there are stringent requirements for both ensuring adequate protection of image privacy and maintaining the integrity and quality of image content. As a pivotal characteristic of data privacy protection strategies in this context, data-level reversibility aims to achieve pixel-by-pixel precise recovery of encrypted or concealed images in the hands of authorized users, meeting the standard of lossless reconstruction. This feature not only satisfies the need for privacy protection but also guarantees the complete reproduction of image information when necessary, making it an indispensable element in constructing efficient and reliable data-level privacy protection solutions.

(2) Data-Level Irreversibility. Data-level irreversibility refers to the process of image protection where pixel-level perfect reconstruction is not pursued, but instead aims to restore an approximate version that is highly similar but not identical to the original image. This strategy is particularly suitable for most scenarios that do not require specific high-precision requirements. It allows for a flexible balance between the accuracy of image reconstruction and other critical attributes of the privacy protection scheme, such as processing efficiency, data security, and robustness. In other words, by moderately sacrificing some precise details of image reconstruction, we can significantly enhance the overall effectiveness and practicality of the privacy protection mechanism.

\subsection{Main Approaches}
\subsubsection{Cryptography}

Cryptography is a technology for encrypting, analyzing, identifying, and confirming information, as well as managing keys. In the field of image privacy protection, cryptography is particularly crucial. It employs highly specialized encryption methods to transform sensitive or private image data into seemingly random and indecipherable noise-like forms, effectively preventing unauthorized access and prying. For instance, chaos encryption stands out due to its unique nonlinear dynamic characteristics. Leveraging the high sensitivity and unpredictability of chaotic systems, it provides robust randomness and complexity guarantees for image encryption. 


\subsubsection{Compressed Sensing}
Compressive sensing (CS) technology utilizes the sparsity of signals and achieves synchronous sampling and compression of signals at an extremely low sampling rate through a single, non-adaptive measurement method, with excellent recovery results. In data-level image privacy protection, compressive sensing technology demonstrates its unique advantages. It can directly compress data during the sampling process and seamlessly integrate encryption technology into this process, ensuring that image data is adequately protected while being compressed, thereby enhancing the security of data during transmission and storage. This effectively reduces data redundancy, significantly improving transmission efficiency.

\subsubsection{Neural Network}

Neural networks are hierarchical structures composed of interconnected nodes, designed to mimic the behavior of the human brain. These nodes communicate through weighted connections and perform a series of computations within the network to process input data and generate outputs. In the realm of image privacy protection, numerous scholars have  employed neural networks, leveraging cutting-edge techniques such as domain transfer and the generation of adversarial examples, to provide a robust layer of privacy protection for image data. These methods effectively obscure sensitive information within images, offering a strong safeguard for the secure sharing and utilization of image data. 


\subsection{Solutions for Robustness}
\subsubsection{Data-Level Reversibility Schemes}

Data-level privacy protection solutions are ideal for a wide range of applications, including medical and nature images. Zhang et al. \cite{10} introduced a new 2D chaotic map by analyzing the conventional Chebyshev map and the Sine map. The 2D-CSM exhibits excellent chaotic behaviors across a broad range of parameters and has continuous chaotic intervals, which makes up for the shortcomings of some traditional chaotic maps with narrow and discrete chaotic intervals. Singh et al. \cite{11} proposed a medical image encryption technique based on biometric features, a parametric all-phase bi-orthogonal sine transform, singular value decomposition, and QR decomposition. The core idea of their approach is to capture the patient's fingerprint and utilize it to generate a key management system. This system generates keys that are derived from the patient's unique biometric information, thereby enhancing the security of the encrypted medical images. Chen et al. \cite{9} proposed an efficient privacy-preserving forensic method with analytical statistical properties, designed to address the problem of camera model identification in a highly efficient and secure manner. They employed scrambling encryption to protect the privacy of image content and noise linear mapping processing to conceal the camera model identity of the queried image. To further ensure the lossless reconstruction of secret images, Puteaux et al. \cite{12} designed a correction method for noise-encrypted images, which introduces a classifier to distinguish between clear and encrypted pixel blocks, and performs blind correction on noise-encrypted images while maintaining the image structure.

In the information security, the introduction of decimals makes the topic of computational accuracy inevitable. The definition and related proofs of the two-parameter fractal sorting matrix \cite{13} expand the fractal theory and solve the limitation of computational accuracy in information security. Based on the idea of Moran construction \cite{14}, an iterative construction method of the two-parameter fractal sorting matrix is given. To further improve the robustness of the algorithm, Liu et al. \cite{17} introduced nonlinear exponential terms and high-order power terms into the 1D cosine chaotic mapping system to increase the dimensionality of the 1D chaotic system. The introduced nonlinear exponential terms and high-order power terms are used as chaotic disturbance sources to disturb the iterative process of cosine chaotic mapping, thereby constructing a new 2D exponential-cosine chaotic system.

Wang et al. \cite{15} designed a chaotic image encryption algorithm based on matrix semi-tensor product and composite keys. By using the keys as the initial values of the mixed linear-nonlinear coupled map lattice system, a chaotic sequence is generated to achieve high security and is suitable for color images. In addition, a better method can simultaneously encrypt gray-scale images, color images, and 3D images \cite{16}. By constructing some internal relationships between graphs and images through a flexible adjacency matrix, the graph data structure is extended to image operations to solve the security defects of classical diffusion structures in image cryptographic systems, such as sequential diffusion paths and fixed-pattern diffusion operators. Yuan et al. \cite{29} proposed a method for encrypting DC and AC coefficients, which achieves a balance between file size expansion, format compatibility, and statistical feature changes.

\subsubsection{Data-Level Irreversibility Schemes}
Zhang et al. \cite{18} used nonlinear operations to encrypt the original image for the first time in the encryption-then-compression scheme based on CS, while achieving low computational complexity and high security in linear operations. Liang et al. \cite{20} devised a reversible multi-level privacy scheme grounded in compressive sensing, which grants differentiated access privileges to various users. Semi-authorized users are limited to obtaining feature vectors within the compressed domain, precluding access to the original data, whereas fully authorized users can reconstruct the original signal with high fidelity. There was work that proposed a novel medical data transmission framework from the perspectives of semi-tensor product compressive sensing and hybrid clouds \cite{21}, which guarantees the efficiency, confidentiality, and verifiability of data transmission. Considering both security and robustness during transmission, Wang et al. \cite{22} proposed an image robust encryption algorithm based on scrambled block compressive sensing. They used discrete wavelet transform (DWT) to sparsely represent the image and adopted scrambled block compressive sensing to conduct measurements, which effectively reduces the amount of data transmission and serves as the first layer of encryption. 

Recognizing the limitations of existing CS-based image encryption algorithms in terms of low reconstruction quality and inadequate security performance, Gong et al. \cite{23} proposed a compression encryption algorithm grounded in chaotic systems. Drawing on the continuous Hopfield neural network model, they crafted a novel four-dimensional chaotic system and optimized its weight parameters to ensure superior dynamic capabilities. Additionally, they devised a fresh fractal curve, inspired by the Hilbert curve, to facilitate efficient image shifting, thereby resolving the aforementioned issues in CS-based image encryption. Gao et al. \cite{24} presented a versatile multi-image hybrid encryption algorithm capable of flexibly encrypting color and grayscale images of various sizes. To achieve efficient compression and encryption of images, Wang et al. \cite{25} designed an Encrypt-then-Lossy Compression (ETLC) scheme based on non-uniform subsampling and a custom deep network. By combining uniform and random sampling, this approach enables arbitrary compression ratios for encrypted images. The lossy reconstruction of decrypted and decompressed images is formulated as a constrained optimization problem, and a tailored deep neural network specifically designed for ETLC was developed to address this issue.

Wang et al. \cite{26} constructed a multi-image encryption algorithm utilizing gyrator transform and multi-resolution singular value decomposition (SVD). In this algorithm, the rotation angle parameter of the gyrator transform can be randomly selected, while the multi-resolution SVD exhibits multi-scale adaptability, resulting in a high level of security for the encryption system. The chaotic sequences are closely related to the plaintext images, enabling the algorithm to effectively resist chosen-plaintext attacks. Subsequently, Qin et al. \cite{27} formulated a JPEG image encryption approach that utilizes an adaptive prediction method for anticipating direct current (DC) coefficients. By integrating prediction errors with random integers, they encrypt the histogram of DC coefficient prediction errors, leading to reduced coding length and ensuring minimal increase in file size. Feng et al. \cite{28} proposed a privacy-preserving image retrieval scheme based on image encryption, which guarantees that encrypted images can still extract certain features for use in retrieval tasks.

\subsection{Solutions for Non-Robustness}
\subsubsection{Data-Level Reversibility Schemes}
Based on cryptography technology, privacy protection methods can not only directly and effectively safeguard the confidentiality of image data, but also maintain its privacy unviolated while ensuring that image data is fully utilized in performing specific tasks. This method not only enhances data security but also promotes the legitimate and compliant use of image information in privacy-sensitive environments. Zhu et al. \cite{30} proposed a novel privacy-preserving mahalanobis distance comparison method to improve the accuracy of medical image retrieval. By combining the Mahalanobis distance-based fuzzy c-means algorithm, accurate and confidential medical image retrieval on encrypted data is achieved. To balance privacy and efficiency, Anju et al. \cite{31} proposed a new encryption and indexing scheme for secure image storage on cloud servers. This encryption involves the substitution and permutation of sub-blocks in each block of each channel of every image using different keys. By using local color layout descriptors, the image encryption achieves almost perfect security. Li et al. \cite{32} presented an adaptive verifiable privacy-preserving medical image retrieval scheme in outsourced clouds. The improved logical chaotic mapping algorithm not only satisfies retrieval tasks while providing robust security for medical image datasets but also minimizes the computational overhead associated with the image encryption process.

Yang et al. \cite{135} designed personal-specific veils for visual face privacy protection. The protected images show significant visual differences from the original images, but can still be recognized by facial recognition models. The scheme not only achieves privacy preservation, but also ensures the implementability of the downstream task (i.e., face recognition).

\subsubsection{Data-Level Irreversibility Schemes}
Data-level irreversible privacy protection strategies significantly focus on leveraging neural network technologies, integrating advanced adversarial perturbation mechanisms with neural network-based encryption techniques, to achieve a high level of privacy protection for image data. Ding et al. \cite{33} proposed a deep learning-based encryption and decryption network to complete the encryption and decryption process of medical images. This scheme applies deep learning technology to the field of image-to-image translation, realizing the encryption process of medical images. This encryption method boasts advantages such as a large key space, one-time padding, and sensitivity to key changes. Zhu et al. \cite{34} introduced a multi-center privacy protection network, a novel framework designed for secure medical image segmentation in multi-center collaboration. It provides a new approach for multi-center collaborative learning, capable of reducing data transmission volume and enhancing data privacy protection.

Wu et al. \cite{35} presented the first data abuse prevention mechanism, which is an input converter for deep learning inference services on the user side. It can remove unnecessary information related to the target deep learning inference services. The input data converted by this method maintains good inference accuracy and makes it difficult to manually or automatically annotate for new model training. Liu et al. \cite{36} developed a new privacy-preserving image classification scheme, which can directly apply classifiers trained in the plaintext domain to classify encrypted images without retraining dedicated classifiers. In addition, the encrypted images can be decrypted back to their original form with high fidelity using a key. Su et al. \cite{37} blurred the visual information of data by generating confusing adversarial perturbations while maintaining the correct predictions of the model for hidden targets. Without modifying the parameters of the application model, it makes it flexible for different scenarios. Fitw et al. \cite{38} encrypted the color channels of frames through a new chaotic image scrambling technique based on sine mapping to ensure end-to-end privacy of frame content.

\subsection{Design Principles}
Tables 1 and 2 respectively provide a breakdown of robust and non-robust solutions for data-level privacy protection. Based on the detailed descriptions and classifications of various schemes mentioned above, we summarize several common design principles for this type of privacy protection methods.

\begin{itemize}
\item {\bfseries{Visual invisibility.}} Data-level privacy protection focuses on processing pixels, and the processed images do not reveal any content information of the original image. Although some specific algorithms may leave weak texture traces visually due to their inherent mechanisms, these traces are not sufficient to reveal the specific content of the original image. Therefore, each method ensures the invisibility of the secret image.

\item {\bfseries{Privacy-sensitive area fixity.}} The delineation of privacy-sensitive areas is strictly based on the personalized privacy needs of image owners and is meticulously categorized. A unique feature of this privacy protection framework is that it assumes that all content in the entire image is considered as privacy-sensitive, meaning that the image owner does not designate any specific regions as non-sensitive, but instead adopts a comprehensive and unified protection strategy. This uniform approach ensures that health information in medical images, strategic secrets in military images, and personal privacy elements in natural images are all included in the same critical protection category. 


\item {\bfseries{Contextual privacy.}} Data-level privacy protection effectively prevents any individual from directly accessing the image data. Even attempts to correlate and analyze the data through other modalities cannot pry into the specifics of the image itself. This strong protection ensures that even in complex contextual environments, sensitive confidential information cannot be inferred by indirect means, thus greatly enhancing the privacy and security level of image data.

\end{itemize}

\begin{table}
\centering

\caption{A Summary of Robustness Solutions in the Data Level}
\scalebox{0.9}{
\begin{tabular}{cccccccccc} 
\toprule
\multirow{2}{*}{Sub-class}         & \multirow{2}{*}{Paper } & \multirow{2}{*}{Year} & \multirow{2}{*}{\makecell[c]{Key \\Technique}} & \multirow{2}{*}{FC } & \multirow{2}{*}{UI } & \multirow{2}{*}{DT } & \multicolumn{2}{l}{\makecell[c]{IF}} & \multirow{2}{*}{Distinctive Feature }  \\ 
\cmidrule{8-9}
                           &                     &                     &                     &                     &                     &                     & G & C                 &                      \\ 
\midrule
\multirow{10}{*}{\makecell[c]{Data-Level\\ Reversibility}  } & \cite{9}     & 2022   & CR   &\ding{51}  & \ding{53}                    & \ding{53}  & \ding{51}  & \ding{51}  & Focus on camera shooting \\
& \cite{10}              & 2024   & CR    & \ding{53}  & \ding{51}    &\ding{53} &\ding{51} &\ding{53}                   &Based on chaotic                      \\
& \cite{11}              & 2021                & CR                                     & \ding{53}                    & \ding{51}                    &\ding{53}                     &\ding{51} &\ding{53}                  &Key based on biometric features \\
                           & \cite{12}              & 2021                & CR                                     & \ding{53}                    & \ding{51}                    &\ding{53}                     &\ding{51} &\ding{53}                  &Focus on noise image encryption \\
                           & \cite{13}              & 2022                & CR                                     & \ding{53}                    & \ding{51}                    &\ding{53}                     &\ding{51} &\ding{53}                   &Based on matrix mapping                     \\
                           & \cite{15}              & 2020                & CR                                     & \ding{53}                    & \ding{51}                    &\ding{53}                     &\ding{51} &\ding{51}                   &Key based on Boolean network                    \\
                           & \cite{16}              & 2024                & CR                                     & \ding{53}                    & \ding{51}                    &\ding{53}                     &\ding{51} &\ding{51}                  &Based on chaotic systems                      \\
                           & \cite{17}              & 2022                & CR                                     & \ding{53}                    & \ding{51}                    &\ding{53}                     &\ding{51} &\ding{53}                   & Based on chaotic systems                       \\
                           & \cite{29}              & 2024                & CR                                     & \ding{51}                    & \ding{51}                    &\ding{53}                     &\ding{51} &\ding{53}                  &\makecell{Balance security and\\ file size increments}                     \\ 
\midrule
\multirow{12}{*}{\makecell{Data-Level\\ Irreversibility}  }        & \cite{18}              & 2021                &CS                     & \ding{51}                    & \ding{51}                    &\ding{53}                     &\ding{51} &\ding{53}                   &High efficiency                      \\
                           & \cite{20}              & 2023                &CS                       &                    \ding{51} & \ding{51}                    & \ding{51}                    &\ding{51}   &                  \ding{53}                     & Multi-level permissions                    \\
                           & \cite{21}              & 2023                &CS &                    \ding{51} & \ding{51}                    &\ding{53}                     &\ding{51}   &                  \ding{53}                     & Certifiable                     \\
                           & \cite{22}              & 2022                & CS                    &                    \ding{51} & \ding{51}                    &\ding{53}                     &\ding{51}   &                  \ding{53}                     & Based on chaotic                     \\
                           & \cite{23}              & 2024                &NN                     &                    \ding{51} & \ding{51}                    &\ding{53}                     &\ding{51}   &                  \ding{51}          &                     Based on chaotic  \\
                           & \cite{24}              & 2023                &NN                     &                    \ding{51} & \ding{51}                    & \ding{53}                    &\ding{51}   &                  \ding{51}  &                     High flexibility \\
                           & \cite{25}              & 2023                &NN                     &                    \ding{51} & \ding{51}                    & \ding{53}                    &\ding{51}   &                  \ding{51}  &       Encrypt first and then compress \\
                           & \cite{26}              & 2020                &CR                     &                    \ding{51} & \ding{51}                    & \ding{53}                    & \ding{51}  &                  \ding{51}  &  Multi image encryption \\
                           & \cite{27}              & 2023                &CR                     &                    \ding{51} &\ding{51}                     & \ding{53}                    &\ding{51}   &                  \ding{53}   &     \makecell[c]{Balance security and\\ file size increments} \\
                           & \cite{28}              & 2024                &CR                     &                    \ding{51} &\ding{51}                     &  \ding{51}                   &\ding{51}   &                  \ding{51}   &    \makecell[c]{Based on encrypted\\ visual transformation} \\

\bottomrule
\end{tabular}}

 \begin{tablenotes}
        \footnotesize
        \item[*] CR: Cryptography; NN: Neural Network; CS: Compressed Sensing; FC: Format Compatibility; UI: Universality of Images; DT: Downstream Task; IF: Image Format; G: Grayscale Image; C: Color Image.  
      \end{tablenotes}
\end{table}

\begin{table}
\centering

\caption{A Summary of Non-Robustness Solutions in the Data Level}
\scalebox{0.9}{
\begin{tabular}{cccccccccc} 
\toprule
\multirow{2}{*}{Sub-class}         & \multirow{2}{*}{Paper } & \multirow{2}{*}{Year} & \multirow{2}{*}{\makecell[c]{Key \\Technique}} & \multirow{2}{*}{FC } & \multirow{2}{*}{UI } & \multirow{2}{*}{DT } & \multicolumn{2}{l}{\makecell[c]{IF}} & \multirow{2}{*}{Distinctive Feature }  \\ 
\cmidrule{8-9}
                           &                     &                     &                     &                     &                     &                     & G & C                 &                      \\ 
\midrule
\multirow{4}{*}{\makecell{Data-Level\\ Reversibility}  } & \cite{30}               & 2023                & CR                   &\ding{53}                   & \ding{51}                    & \ding{51}                   & \ding{51}  & \ding{51}                  & Feature descriptors \\
                           & \cite{31}              & 2024                & CR                   & \ding{53}                    & \ding{51}                    &\ding{51}                     &\ding{51} &\ding{51}                   &Image signature \\
                           & \cite{32}              & 2024                & CR                                     & \ding{53}                    & \ding{51}                    &\ding{51}                     &\ding{51} &\ding{51}                  &                     Image retrieval \\
                           & \cite{135}              & 2024                & NN                                     & \ding{53}                    & \ding{51}                    &\ding{51}                     &\ding{51} &\ding{51}                  &                     Versatility \\   
                        
\midrule
\multirow{6}{*}{\makecell{Data-Level\\ Irreversibility}  }        & \cite{33}  & 2021                &NN                     & \ding{53}                    & \ding{53}                    &\ding{53}                     &\ding{51} &\ding{53}                   & Focus on medical images                     \\
                           & \cite{34}              & 2024                &NN                      &                    \ding{53} & \ding{53}                    & \ding{51}                    &\ding{51}   &                  \ding{53}                     &Focus on medical images                     \\
                           & \cite{35}              & 2021                &NN                     &                    \ding{53} & \ding{51}                    &\ding{51}                     &N\textbackslash A   &                  \ding{51}                     & Combat data abuse                    \\
                           & \cite{36}              & 2024                & NN                 &                    \ding{53} & \ding{51}                    &\ding{51}                     &N\textbackslash A   &                  \ding{51}     & Adversarial perturbations \\
                           & \cite{37}              & 2023                &NN                     &                    \ding{53} & \ding{53}                    &\ding{51}                     &N\textbackslash A   &                  \ding{51}                              & Adversarial perturbations       \\
                           & \cite{38}              & 2023                &NN                    &                    \ding{53} & \ding{51}                    &\ding{53}                     &N\textbackslash A   &                  \ding{51}      &  Focus on surveillance scenarios   \\

\bottomrule
\end{tabular}}

 \begin{tablenotes}
        \footnotesize
        \item[*] CR: Cryptography; NN: Neural Network; CS: Compressed Sensing; FC: Format Compatibility; UI: Universality of Images; DT: Downstream Task; IF: Image Format; G: Grayscale Image; C: Color Image.  
      \end{tablenotes}
\end{table}

\section{CONTENT-LEVEL IMAGE PRIVACY PROTECTION}
This section analyzes in detail the content-level image privacy protection scheme, including its abilitys, main methods, and common design principles. Fig. 5 provides an overview of this stage.

\begin{figure}[h]
  \centering
  \includegraphics[scale=0.4]{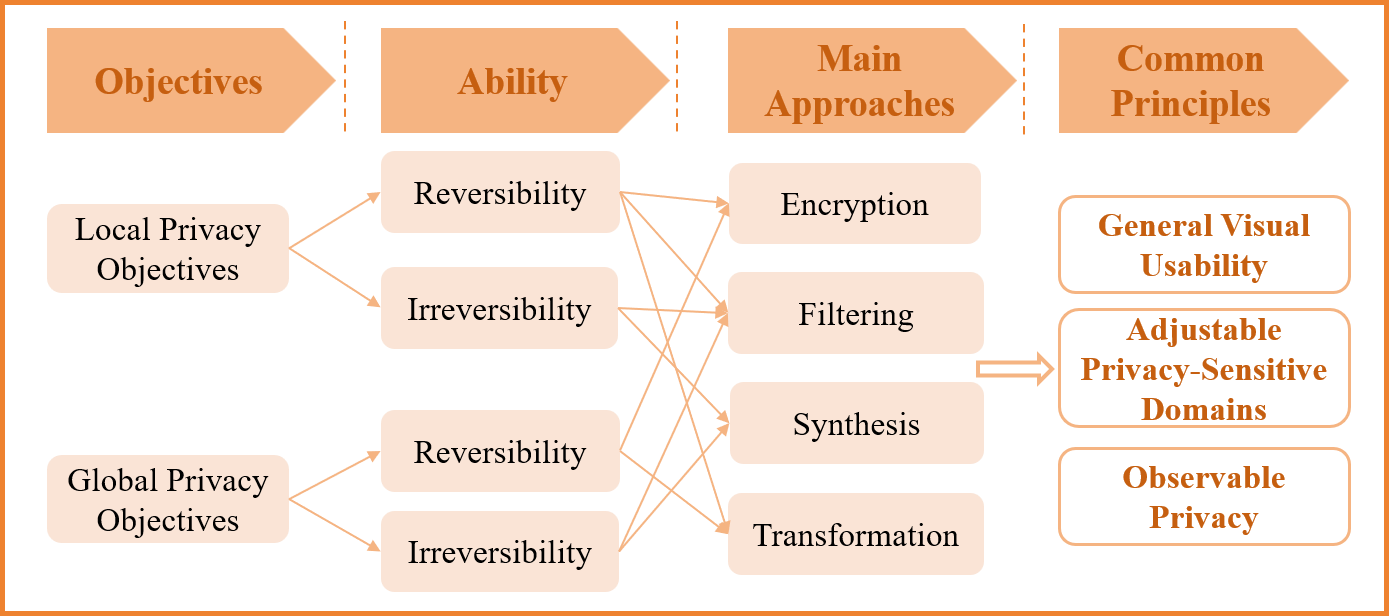}
  \caption{Overview of the content-level image privacy protection.}
\end{figure}

\subsection{Abilitys of Content-Level Privacy Protection}
Content-level image privacy protection technology focuses on the visual content of images as privacy-sensitive areas, ensuring that the images visible to attackers are complete and visually meaningful through the transformation or synthesis of private content, but do not contain any private information. Similar to data-level privacy protection, when discussing content-level image privacy protection, we also pay attention to its two core abilitys, namely reversibility. However, for human eyes, visual information can be recognized as its original information in most cases as long as the content is similar. Therefore, in content-level image privacy protection, we do not need to pursue the complete restoration of each pixel point like data-level privacy protection, but only need to focus on the reversibility of the content. Below, we will elaborate on the two core abilitys of content-level image privacy protection.

(1) Content-Level Reversibility. It implies that the protected image, when reconstructed by the receiver, can be restored to an image identical to the original content. This restoration may not be lossless, but it suffices as long as the content remains consistent. Since visual content can provide us with abundant information, even when there is some loss in content, we can still perceive the original content of the image through clues such as its outlines, colors, or background. However, this method is not suitable for scenarios with high requirements for image accuracy but is more commonly seen in situations where images are only needed for sharing, such as social networks and smart surveillance.

(2) Content-Level Irreversibility. It refers to the inability to restore the protected content into its original or nearly original form. This method employs irreversible technical means to achieve the purpose of privacy protection while still retaining sufficient visual information for use by others or for specific tasks by institutions. The core of this method does not lie in restoring the image but rather in the hope of sharing images without revealing personal privacy, allowing others to see the content he wishes to share or using these images for specific tasks such as face recognition, dataset training, etc.

\subsection{Main Approaches}
\subsubsection{Filtering}
Filtering, as a common strategy in the field of privacy protection, achieves the goal of protecting privacy by reducing the clarity of information in privacy-sensitive areas of images. This method can be mainly subdivided into two categories: blurring and pixelation. Blurring, at its core, utilizes the mathematical tool of the Gaussian function to refine the visual content of an image. Specifically, this process leverages the interrelationships between adjacent pixels to reassign values to each pixel point through weighted averaging, resulting in the smoothing of image content. This treatment not only effectively diminishes the detailed information within the image but also preserves the overall visual coherence, making privacy-sensitive areas visually blurred and difficult to identify.


\subsubsection{Synthesis}
Synthetic methods, as an advanced strategy for privacy protection, focus on modifying or replacing the content of original image information to ensure that the entire image remains visually coherent and consistent in content while protecting privacy. This approach offers unique advantages over filtering techniques. In synthetic methods, the first step is to accurately identify privacy-sensitive areas within the original image. Subsequently, these sensitive areas undergo content modification or replacement through image processing techniques, rather than simply blurring or pixelating them. The modified or replaced portions must seamlessly blend with the rest of the original image to avoid creating any jarring or inconsistent effects. The resulting image, after synthetic processing, exhibits changes in the content of privacy-sensitive areas but overall maintains similar visual characteristics and scene ambiance to the original image. 


\subsubsection{Transformation}
Transformation technology, as a unique means of privacy protection, revolves around thoroughly converting the privacy-sensitive areas of an image into entirely different visual elements from the original content. A notable distinction between transformation technology and synthetic techniques lies in its independence from the specific content of the original image for modification or replacement. Instead, it adopts brand new content as the carrier to achieve the concealment and obscuring of private information. Images processed by transformation technology have their privacy-sensitive regions replaced by designed graphics, patterns, or other such elements that have no direct connection to the original image content, thereby creating a striking visual contrast and difference. This approach not only ensures that private information is effectively protected but also imparts a fresh visual aspect to the image.

\subsubsection{Encryption}
Encryption technology secures the transmission of images by converting them into a visually meaningless and chaotic form, ensuring that only authorized personnel with the correct key can decrypt and access the original image content. This transformation fundamentally protects the privacy of image content, rendering the encrypted image information unreadable by unauthorized individuals, even if they receive it. Unlike schemes in data-level image privacy protection, the encryption strategy here does not involve comprehensively encrypting the entire image as a privacy-sensitive domain. Instead, it adopts a more detailed and flexible approach by encrypting only select pixels within the image. This method, while protecting image privacy, also takes into account the visual usability of the image, ensuring that the encrypted image retains a certain degree of its original visual information and recognizability.

\subsection{Solutions for Local Privacy Objectives}
\subsubsection{Content-Level Reversibility Schemes}
Dou et al. \cite{39} proposed a privacy protection scheme based on image rendering and data hiding. By hiding privacy targets within other data, and performing two rounds of inpainting around the edges of the privacy targets, followed by fusing the inpainted results, they achieved anti-forensic functionality. Liu et al. \cite{40} introduced a complementary embedding encryption strategy that first identifies airport regions and then replaces them with optimally similar regions, embedding the airport images into random locations through a complementary embedding algorithm. Boult et al. \cite{41} presented an encryption scheme for face regions, an application of encryption technology aimed at improving privacy while allowing general surveillance to continue, and enabling full access only with decryption keys maintained by courts or other third parties. Cao et al. \cite{45} proposed a reversible visual transformation algorithm for protecting JPEG image content, where the DC coefficient in each user-selected block is modified, and the information required for its restoration is reversibly hidden in the AC coefficients. Subsequently, the signs of the AC coefficients in the selected blocks are flipped, and the blocks are further scrambled using a key. By embedding the location information of the selected blocks in the transformed image, the original image can be accurately restored when needed. Additionally, Ruchaud et al. \cite{48} focused on balancing privacy protection and comprehensibility while preventing re-identification after common de-anonymization attacks. Their approach automatically adjusts the protection level based on the resolution of regions of interest and demonstrates robustness against de-anonymization attacks. He et al. \cite{19} presented a multi-level image privacy protection scheme based on 2DCS. This scheme utilizes random orthogonal matrices and Arnold scrambling operations to confuse and encrypt different sensitive regions of an image, while 2DCS is responsible for compressing, sampling, and re-encrypting the image.

Yuan et al. \cite{42,46} proposed a framework for protecting image visual privacy in a secure, reversible, highly flexible, and personalized manner. This framework allows for the application of arbitrary regional visual manipulations to regions of interest (ROIs) within an image, while secretly preserving information about the original ROIs within the application segment of visually obfuscated JPEG images. Furthermore, Yuan et al. \cite{43} also designed a privacy-preserving photo sharing architecture that maintains the usability and convenience of online photo sharing activities while ensuring user privacy.

In general, operations such as pixelation and blurring are irreversible, but You et al. \cite{44} proposed a novel reversible facial privacy protection scheme framework based on mosaicking, which blurs facial regions using mosaicking techniques. For low-privilege users, they can perform computer vision tasks on the protected images using the provided classifier. For authorized users, after content restoration, the normal use of facial images will not be affected.

\subsubsection{Content-Level Irreversibility Schemes}
Yu et al. \cite{49} developed a new tool called iPrivacy to automate the privacy setting process during social image sharing. This solution automatically detects privacy-sensitive objects from shared images, identifies their categories, and recognizes their privacy settings. Based on the detection results, it warns the image owner about which objects in the image need protection before sharing and provides recommended privacy settings. Sun et al. \cite{52} and Zhou et al. \cite{53} proposed separate privacy protection schemes for face pixelation, both of which consider consistency when applied to videos. Orekondy et al. \cite{129} employed automated methods to edit private and sensitive information, thereby protecting individual privacy. To prevent blurred images from affecting downstream task performance, Jiang et al. \cite{51} proposed a novel artifact-removal blurring privacy protection method that utilizes a deep neural network architecture to generate blurred faces, effectively hiding facial privacy information while suppressing detection artifacts. Shvai et al. \cite{50} implemented the blurring anonymization of rectangular image blocks containing sensitive information, allowing existing classifiers to maintain decision invariance on anonymized images without any modifications to the classification models.

Morris et al. \cite{54} defined a novel fine-grained location-aware multi-party image access control mechanism, which breaks the limitation of existing privacy protection only for photo owners and their friends, and provides equal privacy protection for each identifiable individual in the photo, rather than for the photo owner and their friends. Once the user's identity is confirmed, and according to the user's privacy policy, if the location of the photo is considered sensitive, the user's face will be replaced with a synthetic face. Yang et al. \cite{55} developed Digital Mask (DM), a new technology based on real-time 3D reconstruction and deep learning algorithms. DM takes the original image as input and outputs a reconstructed image containing disease information, while discarding as much of the patient's identity as possible, highlighting fine-grained eye reconstruction, and the reconstructed image by DM cannot be converted back to the original image because most of the information required to reconstruct the original attributes has been discarded. In addition, Brkic et al. \cite{56} proposed a human body and face image de-identification model. Given the segmentation of the human body, the model can generate a synthetic human image with an alternative appearance that looks natural and conforms to the segmentation contour.

\subsection{Solutions for Global Privacy Objectives}
\subsubsection{Content-Level Reversibility Schemes}
Thumbnail-preserving encryption (TPE) technology that balances image usability and privacy by blurring details within thumbnail blocks while preserving contours has emerged as a key technology for image privacy protection \cite{57,58,59,66,67,74,60,63,64,61,62,65}. The first approach to this technology was proposed by Wright et al. \cite{57}, which maintain the average pixel within a block unchanged by only changing the positions of pixels within a sub-block without altering any pixel values, thereby achieving consistency between the encrypted image and the thumbnail of the original image. Tajik et al. \cite{58} introduced a precise thumbnail-preserving encryption scheme and demonstrated that the encrypted ciphertext image under this scheme does not reveal any information related to the original image beyond the thumbnail. Zhao et al. \cite{59} further extended the TPE scheme by proposing a three-pixel-precise TPE scheme, known as TPE2. This scheme realizes precise TPE as a group of three pixels. Additionally, two visually meaningful encryption schemes \cite{66,67} were also proposed, both of which exhibit good recovery quality. 

Ye et al. \cite{103} proposed a joint learning reversible anonymity framework that can reversibly generate full-body anonymous images. An identity-specific encryption-decryption architecture is introduced to randomly encrypt anonymous images based on specific keys assigned to each identity. To completely hide image content while still maintaining visual significance in the transmitted image, Lu et al. \cite{68} designed a high-capacity reversible steganography network for image steganography, treating steganography and the recovery of hidden images as a pair of inverse problems on image domain transformations. Then, the forward and backward propagation operations of a single reversible network are introduced to embed and extract images. To improve the robustness of the embedding network, Ying et al. \cite{69} introduced a method based on generative deep networks to hide images within images, ensuring that high-quality images can be extracted from destructive composite images by constructing a real-world attack simulator. To achieve multi-image steganography, Guan et al. \cite{70} developed a reversible hidding neural network, innovatively modeling image hiding and display as its forward and backward processes, making them fully coupled and reversible. Through an importance map module, the current image hiding is guided based on previous image hiding results, allowing for multiple cascades as needed to achieve the hiding of multiple images. 

The innovative aspect of the aforementioned schemes lies in their focus on lossless spatial images as carriers for data hiding. However, this characteristic limits their direct application in widely used JPEG lossy format images. In view of this, Yang et al. \cite{71} pioneeringly proposed a solution that successfully embedded image-level covert information of the same scale into color JPEG images. This scheme fully exploits fine-grained DCT representations conducive to hiding and recovery, providing some inspiration for further research on hiding image-level messages in JPEG carrier images and extending the types of secret images to JPEG images.

For DNN-based steganography schemes, the steganographic tool must be transmitted to both the sender and the receiver for secret embedding and recovery, inevitably increasing the overall load of data transmission. Therefore, Li et al. \cite{73} attempted to solve the DNN model steganography problem with a purified and unified steganography network that can simultaneously hide the secret encoding and decoding networks within a purified network. It can perform ordinary machine learning tasks without being detected. The hidden steganography network can be triggered from the purified network using specific keys possessed by the sender or receiver.

Smartphones can easily access offline-to-online information by scanning QR codes, but as the demand for experience improves, the appearance of QR codes also needs to be more diverse. Jia et al. \cite{72} proposed a new model for invisible information hiding in display/print camera scenarios. This model generates marks that are invisible to the human eye but detectable by the positioning network through joint training of the encoder and positioning network. Such QR codes can seamlessly blend into visually appealing images in a concealed form, without affecting the aesthetic value of the image while ensuring that the QR code is effectively scanned and recognized.

For facial privacy enhancement, Yuan et al. \cite{75} proposed a new paradigm to protect facial privacy through a reversible image obfuscation framework. This paradigm utilizes a flow-based model to transform face images into a blurred form relative to a pre-blurred template in a reversible and secure manner. Furthermore, through a specially designed key mechanism, the recovery of privacy-protected images is protected, and only entities with the correct key can access the correct image recovery.

Wu et al. \cite{76} designed a multi-functional privacy enhancement system that runs in real-time at the frontend of Video Streaming and Analytics. By introducing CycleGAN \cite{77}, the sysytem performs privacy enhancement transformations and securely enables bidirectional (reversible) transformations only for authorized parties. It utilizes visual style transfer to achieve an adjustable balance between privacy and intelligibility. Ciftci et al. \cite{78} proposed two fully reversible privacy protection schemes implemented within the JPEG architecture. In both schemes, privacy protection is achieved through the use of fake colors. One is applicable to other privacy protection filters, and the other is specific to fake colors. Both schemes support lossless and lossy modes, where the former allows the original unprotected content to be fully extracted, while the latter restricts file size while maintaining intelligibility. Wu et al. \cite{47} can visually alter a secret image by imitating an arbitrarily chosen reference image and precisely recover it from the transformed image when needed.

For medical image privacy protection \cite{79,80,146}, Wen et al. \cite{79} constructed a watermark-cycle consistency network for secure sharing of medical images in telemedicine, converting secret images into other meaningful images for transmission to reduce suspicions from attackers. This network is unique in that only the paired CycleGAN can recover the correct medical image. Networks that have not been trained with the watermark can also recover meaningful images, achieving the purpose of identity confusion. Therefore, the proposed model offers high security in medical image privacy protection. Ping et al. \cite{80} proposed a novel medical image hiding method that simultaneously considers the protection of patients' diagnostic information and corresponding medical images. Within this framework, patients' diagnostic information and medical images can be hidden within a target image for secure transmission, and authorized medical staff can recover the medical images and diagnostic information without distortion.

\subsubsection{Content-Level Irreversibility Schemes}
Ryoo et al. \cite{83} introduced the inverse super-resolution paradigm to anonymize videos for privacy protection through extremely low resolution. Specifically, the learning transformation is performed from a set of high-resolution images, so that the generated low-resolution images retain a comparable amount of information as the high-resolution images. Yuan et al. \cite{84} proposed a novel blurred face privacy protection recognition paradigm based on a dedicated feature compensation mechanism, which subtly balances the delicate relationship between privacy protection and recognition utility. Its core lies in constructing an efficient client-server architecture, where the client is only responsible for securely transmitting blurred images to the server, and the server relies on a pre-trained advanced model complemented by a set of supplementary functional modules that pose no privacy leakage risk to achieve accurate identity recognition. The uniqueness of this framework lies in its ability to not only ensure that the accuracy of face recognition is not significantly affected but also effectively protect the original clarity of individual facial features from being easily revealed, thereby building a solid bridge between privacy protection and maintaining recognition functionality. Ye et al. \cite{86} developed a protective perturbation generator to protect user privacy in deep learning-based mobile image recognition applications. The perturbation generator can effectively blur input images without affecting the prediction accuracy of the target image recognition model, thereby preventing privacy leaks.

Hassan et al. \cite{81} proposed an automatic “cartoonization” transformation to enhance privacy in live broadcasts and first-person images. Just like animated films abstract away the details of the real world to convey only the most important semantic elements, cartoonization can blur private details in videos while still retaining the overall “story”. The parameters of the algorithm can be adjusted to control the aggressiveness of the transformation. Additionally, Erdélyi et al. \cite{82} converted original frames into abstract cartoon frames and removed privacy-revealing details within the abstract cartoon frames. Privacy protection and practical effects can be achieved by adjusting the specific settings of the cartoon filter. Aprilpyone et al. \cite{85} proposed three image transformation algorithms based on encryption ideas. By applying these images processed with encryption techniques to the adversarial training process, the constructed model can not only efficiently classify images in a protected state but also significantly enhance data security. Notably, these algorithms achieve effective masking or deletion of sensitive visual information in images while protecting image content.

\subsection{Design Principles}
Tables 3 and 4 respectively provide a breakdown of solutions for privacy-sensitive domains in content-level privacy protection at the local and global levels. Based on the detailed descriptions and classifications of various schemes mentioned above, we summarize several common design principles for this type of privacy protection methods.

%
%

\begin{table}
\centering

\caption{A Summary of Local Privacy Objectives Solutions in the Content Level}
\scalebox{0.9}{
\begin{tabular}{cccccccccc} 
\toprule
\multirow{2}{*}{Sub-class}         & \multirow{2}{*}{Paper } & \multirow{2}{*}{Year} & \multirow{2}{*}{Key Technique} & \multirow{2}{*}{VU } & \multirow{2}{*}{AP } & \multirow{2}{*}{CI } & \multicolumn{2}{l}{\makecell[c]{PC}} & \multirow{2}{*}{Distinctive Feature }  \\ 
\cmidrule{8-9}
                           &                     &                     &                     &                     &                     &                     & T & I                 &                      \\ 
\midrule
\multirow{10}{*}{\makecell[c]{Content-Level\\ Reversibility}  } & \cite{39}               & 2021                & Encryption                   &\CIRCLE                   & \CIRCLE                    & \Circle                   & \ding{53}  & \ding{51}                  & Inpainting \\
                           & \cite{40}              & 2023                & Encryption                    &\CIRCLE                   & \CIRCLE                    & \Circle                                     &\ding{53} &\ding{51}                   &Similar position replacement                     \\
                           & \cite{41}              & 2005                & Encryption                                      & \LEFTcircle                    & \Circle                    &\CIRCLE                     &\ding{53} &\ding{51}                  &         Focus on surveillance videos \\
                           & \cite{45}              & 2020                & Encryption                                      & \LEFTcircle                    & \Circle                    &\CIRCLE                     &\ding{53} &\ding{51}                  &                     Format compatibility \\
                           & \cite{48}              & 2017                & Encryption                                      & \LEFTcircle                    & \LEFTcircle                    &\CIRCLE                    &\ding{53} &\ding{51}                   & Focus on surveillance videos                    \\
                           & \cite{19}              & 2024                & Encryption                                      & \LEFTcircle                    & \Circle                    &\CIRCLE                     &\ding{53} &\ding{51}                   &  Focus on social networks                    \\
                           & \cite{42}              & 2017                & Transformation                                     & \CIRCLE                    & \LEFTcircle                    &\LEFTcircle                     &\ding{53} &\ding{51}                  & Generality                      \\
                           & \cite{46}              & 2015                & Transformation                                                                         & \CIRCLE                    & \CIRCLE                    &\LEFTcircle                     &\ding{53} &\ding{51}                   & Generality                       \\
                           & \cite{43}              & 2015                & Transformation                                     & \LEFTcircle                    & \Circle                    &\CIRCLE                     &\ding{53} &\ding{51}                   & Format compatibility                      \\
                           & \cite{44}              & 2021                & Filtering                                     & \LEFTcircle                     & \LEFTcircle                     &\CIRCLE                     &\ding{53} &\ding{51}                  &Perform visual tasks                      \\ 
\midrule
\multirow{8}{*}{\makecell{Content-Level\\ Irreversibility}  }        & \cite{49}              & 2017                &Filtering                     & \LEFTcircle                    & \LEFTcircle            &\CIRCLE                     &\ding{51} &\ding{51}                   &\makecell{Automatically ide-\\ntify sensitive areas}        \\
                           & \cite{50}              & 2023                &Filtering & \LEFTcircle                    & \LEFTcircle            &\CIRCLE                                      &\ding{53}   &                  \ding{51}                     &Perform classification tasks                    \\
                           & \cite{51}              & 2023                &Filtering                     &                     \LEFTcircle                    & \LEFTcircle            &\CIRCLE                                      &\ding{53}   &                  \ding{51}         & Suppress artifact detection  \\
                           & \cite{52}              & 2022                & Filtering    &   \Circle   & \Circle                    &\CIRCLE                     &\ding{53}   &                  \ding{51}         & Focus on surveillance videos                    \\
                           & \cite{53}              & 2021                &Filtering                     &                    \Circle & \Circle                    &\CIRCLE                     &\ding{53}   &                  \ding{51}         & Focus on surveillance videos  \\
                           & \cite{54}              & 2023                &Synthesis                     &                    \Circle & \CIRCLE                   & \Circle                   &\ding{53}   &                  \ding{51}  &                     Geographic location privacy \\
                           & \cite{55}              & 2022                &Synthesis                     &                    \LEFTcircle & \LEFTcircle                    & \CIRCLE                    &\ding{53}   &                  \ding{51}  &                     Perform medical diagnosis \\
                           & \cite{56}              & 2017                &Synthesis                     &                    \CIRCLE & \CIRCLE                    & \LEFTcircle                    & \ding{53}  &                  \ding{51}  &                     Authenticity \\

\bottomrule
\end{tabular}}

 \begin{tablenotes}
        \footnotesize
        \item[*] VU: Visual Usability of Secret Images; AP: The Authenticity of the Privacy Section; CI: Privacy Content can be Inferred; PC: Privacy Content; T: Text; I: Image; \Circle: Refuse; \LEFTcircle: Commonly; \CIRCLE: No Problem.  
      \end{tablenotes}
\end{table}

\begin{itemize}
\item {\bfseries{General visual usability.}} Usability is a crucial consideration during image transmission. Images that resemble noise or lack clear meaning tend to be misjudged as encrypted images, inadvertently increasing the risk of being targeted and attempted to be cracked by malicious attackers. In contrast, content-level privacy protection strategies balance the need for privacy and security. They not only protect the core information of the image from being easily leaked but also retain some of the original content of the image, providing a certain level of visual information feedback. This allows the recipient to perceive the basic information outline of the image, thereby protecting sensitive information while maintaining the basic usability and recognizability of the image.

\item {\bfseries{Adjustable privacy-sensitive domains.}} Given the diverse privacy preferences and needs of each user, the design of privacy protection schemes must demonstrate a high degree of flexibility to precisely match and protect the specific content that users are concerned about, rather than adopting a one-size-fits-all comprehensive hiding strategy. In addition, there are also corresponding requirements in special scenarios, such as social networks and cloud storage. Furthermore, in specific application scenarios such as social networks and cloud storage, privacy protection schemes need to further adapt to the unique requirements of the scenarios. For example, in social networks, users may wish to share snippets of their lives while protecting personal identities or location information.

\item {\bfseries{Observable privacy.}} The most direct and significant way to identify the privacy content of an image is through human observation. Therefore, the core objective of privacy protection technology is to use specific technical means to hide or blur these sensitive information, making it difficult for unauthorized human eyes to directly explore the privacy content. Content-level privacy protection methods focus precisely on this challenge, aiming to address privacy leaks caused by the intuitive observation of image content by human eyes, i.e., protecting observable privacy content.

\end{itemize}

\begin{table}
\centering

\caption{A Summary of Global Privacy Objectives Solutions in the Content Level}
\scalebox{0.9}{
\begin{tabular}{cccccccccc} 

\toprule
\multirow{2}{*}{Sub-class}         & \multirow{2}{*}{Paper } & \multirow{2}{*}{Year} & \multirow{2}{*}{Key Technique} & \multirow{2}{*}{VU } & \multirow{2}{*}{AP } & \multirow{2}{*}{CI } & \multicolumn{2}{l}{\makecell[c]{PC}} & \multirow{2}{*}{Distinctive Feature }  \\ 
\cmidrule{8-9}
                           &                     &                     &                     &                     &                     &                     & T & I                 &                      \\ 
\midrule
\multirow{17}{*}{\makecell[c]{Content-Level\\ Reversibility}  } 
                           & \cite{74}              & 2023                & Encryption                                      &\LEFTcircle                  &\LEFTcircle                    & \LEFTcircle                   & \ding{53}  & \ding{51}                  &  Generality               \\
                           & \cite{60}              & 2023                & Encryption                                    &\LEFTcircle                  &\LEFTcircle                    & \LEFTcircle                   & \ding{53}  & \ding{51}                   & Format compatibility \\
                           & \cite{61}              & 2022                & Encryption                                                                         &\LEFTcircle                  &\LEFTcircle                    & \LEFTcircle                   & \ding{53}  & \ding{51}                    & High fidelity                       \\
                           & \cite{62}              & 2023                & Encryption                                     &\LEFTcircle                  &\LEFTcircle                    & \LEFTcircle                   & \ding{53}  & \ding{51}             & High efficiency            \\
                           & \cite{63}              & 2024                & Encryption                                     &\LEFTcircle                  &\LEFTcircle                    & \LEFTcircle                   & \ding{53}  & \ding{51}                  &Non-destructive recovery             \\ 
                           & \cite{103}              & 2024                &Encryption                                     &\LEFTcircle                  &\LEFTcircle                    & \LEFTcircle                   & \ding{53}  & \ding{51}                  &Prevent re-identification           \\ 
                           & \cite{69}              & 2021                & Transformation                                     & \Circle                     & \CIRCLE                    &\Circle                      &\ding{53} &\ding{51}                  &Robustness                     \\ 
                           & \cite{70}              & 2023                & Transformation                                     & \Circle                     & \CIRCLE                    &\Circle                      &\ding{53} &\ding{51}                  &High-capacity                     \\ 
                           & \cite{71}              & 2023                & Transformation                                     & \Circle                     & \CIRCLE                    &\Circle                      &\ding{53} &\ding{51}                  &Format compatibility             \\ 
                           & \cite{73}              & 2024                & Transformation                                     & \Circle                     & \CIRCLE                    &\Circle                     &\ding{53} &\ding{51}                  &Network Model Privacy \\ 
                           & \cite{72}              & 2022                & Transformation                                     & \Circle                     & \CIRCLE                    &\Circle                      &\ding{53} &\ding{51}                  &Automatic correction \\ 
                           & \cite{75}              & 2024                & Transformation                                     & \LEFTcircle                     & \LEFTcircle                     &\Circle                     &\ding{53} &\ding{51}                  &Focus on facial images       \\ 
                           & \cite{76}              & 2021                & Transformation                                     & \CIRCLE                     & \LEFTcircle                     &\CIRCLE                     &\ding{53} &\ding{51}                  &Cartoonization               \\ 
                           & \cite{78}              & 2018                & Transformation                                     & \LEFTcircle                     & \LEFTcircle                     &\LEFTcircle                     &\ding{53} &\ding{51}                  &Colorization     \\ 
                           & \cite{47}              & 2020                & Transformation                                     & \Circle                     & \CIRCLE                     &\Circle                     &\ding{53} &\ding{51}                  &Fixed cover image                \\ 
                           & \cite{79}              & 2024                & Transformation                                     & \Circle                     & \LEFTcircle                     &\Circle                     &\ding{51} &\ding{51}                  &Focus on medical images                     \\ 
\midrule

\multirow{6}{*}{\makecell{Content-Level\\ Irreversibility}  }        & \cite{83}              & 2017                &Filtering                     & \LEFTcircle                    & \LEFTcircle            &\LEFTcircle                      &\ding{53} &\ding{51}                   &Focus on video surveillance        \\
                           & \cite{84}              & 2024                &Filtering & \LEFTcircle                    & \LEFTcircle            &\LEFTcircle                                       &\ding{53}   &                  \ding{51}                     &Focus on facial images                     \\
                           & \cite{86}              & 2022                &Filtering                     &                     \Circle                    & \Circle            &\Circle                                      &\ding{53}   &                  \ding{51}         & Perform recognition tasks \\
                           & \cite{81}              & 2017                & Synthesis    &   \CIRCLE   & \LEFTcircle                    &\LEFTcircle                     &\ding{53}   &                  \ding{51}         & Cartoonization \\
                           & \cite{82}              & 2014                &Synthesis                     &                    \CIRCLE & \LEFTcircle                    &\LEFTcircle                     &\ding{53}   &                  \ding{51}         & Cartoonization  \\
                           & \cite{85}              & 2021                &Synthesis                     &                    \LEFTcircle & \LEFTcircle                   & \LEFTcircle                  &\ding{53}   &                  \ding{51}  &                     Perform classification tasks \\

\bottomrule
\end{tabular}}

 \begin{tablenotes}
        \footnotesize
        \item[*] VU: Visual Usability of Secret Images; AP: The Authenticity of the Privacy Section; CI: Privacy Content can be Inferred; PC: Privacy Content; T: Text; I: Image; \Circle: Refuse; \LEFTcircle: Commonly; \CIRCLE: No Problem.  
      \end{tablenotes}
\end{table}

\section{FEATURE-LEVEL IMAGE PRIVACY PROTECTION}
This section provides a detailed analysis of feature level image privacy protection schemes, including abilitys, main approaches, and common design principles. Fig. 6 provides an overview of this stage.

\begin{figure}[h]
  \centering
  \includegraphics[scale=0.4]{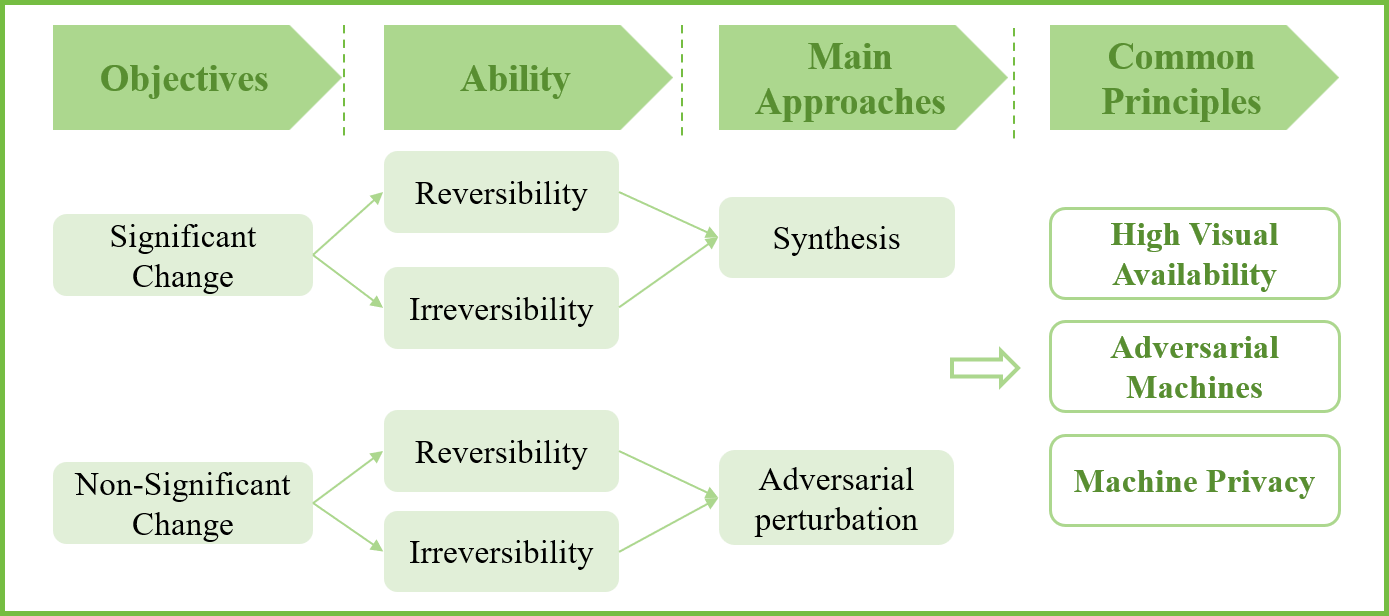}
  \caption{Overview of the feature-level image privacy protection.}
\end{figure}

\subsection{Abilitys of Feature-Level Privacy Protection}
Feature-level privacy protection refers to the ability to use privacy processing techniques to make it difficult for machine learning models or intelligent systems to accurately recognize or make it difficult to misrecognize sensitive high-level feature information contained in images, aiming to balance the needs between technological application and personal privacy protection. These features are subtle image characteristics that are difficult for human vision to perceive but can be keenly captured by machine learning models. Similar to data-level and content-level privacy protection, when exploring image processing strategies, we also need to carefully consider the reversibility of image information. However, at the feature-level, this reversibility transcends the boundaries of visual recognition by human eyes and delves into more refined and complex data dimensions. The following will elaborate on these two abilitys in detail.

(1) Feature-Level Reversibility. Feature-level reversibility refers to the ability to protect the high-level feature information in images from being easily recognized by machines while retaining the capability to restore such feature information. Specifically, after images undergo feature-level privacy processing, although their key features are deleted or obfuscated, making it difficult for ordinary machine learning models or intelligent systems to directly identify sensitive information in the original images, these processed high-level features can be restored to their original state losslessly upon request by authorized users through the application of specific restoration algorithms, enabling machines to re-recognize and analyze the true content of the images. 


(2) Feature-Level Irreversibility. Feature-level irreversibility refers to the fact that after the image is processed at the feature level, the high-level feature information contained in it will be completely erased or converted into an irreversible form. This processing method ensures that any attempt to restore image features will fail, thereby greatly enhancing the strength and security of privacy protection. By adopting feature-level irreversibility technology, users can enjoy unprecedented privacy security. In daily life, especially in the face of ubiquitous video surveillance, people often worry that their personal information and behavior patterns may be captured and abused without authorization. However, these concerns will be greatly alleviated through feature-level irreversible privacy processing, as even if the images are captured, the critical feature information within them has been permanently removed, rendering them unusable for tasks such as identity recognition.

\subsection{Main Approaches}
\subsubsection{Adversarial Perturbation}
Adversarial perturbation aims to embed imperceptible noise or disturbance into natural images to induce deep neural network models to produce incorrect classification or label estimation with high probability \cite{87}. This technology exploits the weakness of neural networks' sensitivity to input data, which can significantly affect the output results of the model even when there are only subtle and imperceptible changes in the input data. This method is not only regarded as a cutting-edge defensive strategy against machine recognition and exploitation, but also one of the most core and mainstream technical means of feature-level privacy protection.

\subsubsection{Synthetic}
Synthetic technology involves the refined elimination or rearrangement and combination of privacy features in images to be protected, thereby generating an image with privacy protection achieved at the feature level. This method is significantly different from the synthesis strategy in content-level privacy protection. It highly relies on advanced generative model technology and focuses more on the processing of the intrinsic features of images, rather than simple content replacement or masking. This ensures that while maintaining the coherence of basic image information, sensitive or private feature information is effectively concealed.

\subsection{Solutions for Significant Change }
\subsubsection{Feature-Level Reversibility Schemes}
With the widespread adoption of deep learning technologies such as deepfake and their potential risks of misuse, strengthening facial feature anonymization and deanonymization techniques has emerged as a critical task in safeguarding personal privacy. Cao et al. \cite{88} introduced user-specific passwords and adjustable parameters to control the direction and extent of identity alteration, ensuring that while the input identity is transformed, other attributes remain similar. Pan et al. \cite{89} designed a reversible privacy protection framework for facial images based on a conditional encoder-decoder architecture. By selecting an input image, a password, and a multi-factor combination, target facial attributes can be generated. During the recovery phase, entering the correct inverse password and the selected multi-factor combination enables the reconstruction of a facial image close to the original. However, receiving an incorrect password or multi-factor combination results in the generation of another facial image with a different identity, thereby enhancing the security of the scheme. Proenca et al. \cite{90} proposed a reversible face de-identification method known as UU-Net. In this work, a reversible de-identification process is described, where two consecutive U-Nets are trained: one for de-identification and the second for reversing the de-identification process. As a result, the first U-Net generates privacy-enhanced video data, facilitating video analysis and data sharing, while the second U-Net remains private and can be accessed by authorities when the original data is required. Li et al. \cite{91} presented a reversible and diversified de-identification with a latent encryptor, which achieves different levels of anonymization using distinct passwords. The true identity can only be decrypted with the correct password; otherwise, the system generates another de-identified face to preserve privacy.

To comprehensively consider the anonymization of multiple biometric attributes, Zhang et al. \cite{92} designed a reversible privacy protection scheme targeting various facial attributes, ensuring both the retention of face verification utility and the protection of various attribute privacy along with the recovery of the original image. Furthermore, in cases where an incorrect password is provided, the visually realistic images with machine classification errors still maintain their authenticity, making it impossible for attackers to determine whether the recovery is correct. He et al. \cite{93} introduced a diffusion model for face privacy protection, ensuring that visual privacy is not compromised while still enabling downstream computer vision tasks such as face detection. Wen et al. \cite{104} proposed a modular architecture for reversible face video de-identification. This framework comprises two modules: the de-identification process, which provides a protective mask for identity information, and the recovery process, which removes the protective mask only when the correct key is provided. Both modules are based on identity disentanglement networks and guided by a crucial motion flow module.

\subsubsection{Feature-Level Irreversibility Schemes}
To naturally obfuscate facial identities \cite{136,137} while protecting privacy in a controllable and measurable manner, Li et al. \cite{94} designed a renowned face de-identification method known as AnonymousNet. AnonymousNet employs a multitude of attribute classifiers to identify facial aspects that require obfuscation. Subsequently, it utilizes a GAN model to generate alternative faces with a natural appearance and then maps these generated faces back to the original ones. Finally, it adds adversarial noise \cite{95}, providing an additional layer of privacy enhancement to the entire framework. Despite being built upon different formal privacy schemes, the authors did not discuss whether AnonymousNet also offers privacy guarantees. Maximov et al. \cite{96} applied Conditional GANs to a Conditional Identity Anonymization GAN model, utilizing an encoder-decoder-type generator network trained in an adversarial manner to achieve a face swapping process. For anonymization, the model first employs landmark techniques to locate and mask the facial region in the input image, and then replaces the masked region with an artificially generated face of an alternative identity. During the rendering process, the output identity is determined by a one-hot encoding vector input into the model's latent space.

To achieve de-identification while preserving facial attributes, SF-GAN was proposed \cite{97}. This model aims to effectively implement face de-identification by generating visually plausible images while retaining as much facial attribute information from the original image as possible, such as facial expressions, gender, hairstyle, and whether glasses are worn. By adopting shallow and deep attribute information, and based on the uniqueness of each type of facial attribute information, different processing strategies are applied to different facial attributes. In this controllable manner, the disentanglement metrics are appropriately incorporated into the objective function, greatly ensuring the effectiveness of attribute preservation.

Chen et al. \cite{98} defined Perceptual Indistinguishability (PI) and proposed a privacy protection mechanism called PI-Net, which achieves image obfuscation under the guarantee of PI. PI refers to the indistinguishability result where it is difficult for an attacker to infer the original image when viewing the anonymized image, while maintaining perceptual similarity, thereby protecting the privacy of image content. On the other hand, PI ensures high data utility (for detection, classification, tracking, etc.). PI-Net performs anonymization on selected semantic attributes, and its architecture generates realistic faces with selected attribute preservation.

A privacy enhancement technique built around Generative Adversarial Networks, called PrivacyNet \cite{99}, was proposed in 2020. PrivacyNet is a semi-adversarial network based on GANs that modifies the input face images such that they can be used for matching purposes by face matchers but cannot be reliably used by attribute classifiers. Furthermore, PrivacyNet allows individuals to select specific attributes (e.g., age and race) that must be obfuscated in the input face images, while allowing the extraction of other types of attributes (e.g., gender). This provides fine-grained control over the leakage of private information in images. Wang et al. \cite{100} presented a method for blurring images that inverts information unrelated to the target privacy attributes or maximizes the uncertainty of attributes. This enables the image to be sufficiently altered to introduce false information or minimize the informational content of attributes, while maintaining the overall appearance of the original input image.

Zhang et al. \cite{101} developed a face privacy protection scheme based on virtual identity transformation, which exhibits strong privacy protection capabilities and high recognizability. Each user possesses a specific identity mask, ensuring that only the identity features extracted from the user's face can approximate a given virtual identity. Kim et al. \cite{105} introduced a client/server privacy-preserving network in the context of multi-center medical image analysis. This method is based on adversarial learning, where medical images are encoded to obfuscate patient identities while preserving sufficient information for specific medical tasks. 


\subsection{Solutions for Non-Significant Change }
\subsubsection{Feature-Level Reversibility Schemes}
Ye et al. \cite{102} proposed a novel gender privacy protection scheme aimed at enhancing gender privacy while supporting reversibility. They constructed an iterative adversarial perturbation method to prevent unauthorized classifiers from recognizing gender, simultaneously concealing gender-related attributes from the classifiers to prevent gender inference from these attributes, thereby enhancing gender privacy. Chen et al. \cite{106} presented a privacy protection algorithm based on visible watermark perturbation. This algorithm utilizes a proposed saliency map fusion algorithm to obtain the watermark embedding region of the image and an adaptive memory harmony search algorithm to optimize the position and transparency of the embedded watermark perturbation. The image classification model cannot correctly recognize images with embedded watermark perturbations. However, doctors with the key can restore these images without distortion, enabling the classification model to accurately recognize them. Zhou et al. \cite{107} introduced an adversarial facial image thumbnail-preserving transformation scheme that ensures consistent thumbnails of corresponding images before and after transformation. Regardless of the size of the thumbnail block, this scheme can prevent face recognition models from identifying the identities of faces in thumbnail images, successfully striking a balance between privacy and usability. Furthermore, in 2023, Zhang et al. \cite{108} made the first attempt to model adversarial attacks and recovery as a unified task. With the support of joint dynamic training, their scheme enables recoverable adversarial samples to maximize attack capabilities. Through a designed dimensionality reducer, the perturbation distribution is optimized, further enhancing recoverability.

\subsubsection{Feature-Level Irreversibility Schemes}
Yang et al. \cite{109} proposed an iterative method for target identity protection through face encryption for black-box face recognition systems. This approach generates adversarial identity masks that cover facial images, concealing the original identity without sacrificing visual quality. Zhong et al. \cite{110} introduced a privacy mask for all face images of an individual, akin to a customized invisibility cloak. This cloak can be applied to all images of ordinary users to prevent malicious face recognition systems from exposing their identities. To further enhance generalization capabilities, Liu et al. \cite{125} proposed a gradient accumulation method that aggregates multiple mini-batch gradients into a one-step iteration gradient, strengthening gradient stability. Additionally, a new scene text black-box perturbation generation method was proposed by Xu et al. \cite{114}, which achieves text-based privacy protection with minimal pixel perturbations under a black-box setting. Liang et al. \cite{115} leveraged the issue of adversarial examples to protect image privacy from detectors based on deep neural networks. Chatzikyriakidis et al. \cite{116} introduced a method that penalizes fast gradient values, operating in the image spatial domain and generating adversarial examples for face de-identification that resemble the original facial images. Cherepanova et al. \cite{117} addressed the privacy issue of capturing images on OSNs to build face databases and proposed the first tool to thwart commercial face APIs by adding adversarial perturbations to images before uploading them to OSNs to evade face recognition. Chhabra et al. \cite{118} designed a novel algorithm based on the concept of adversarial noise addition, which anonymizes selected attributes (or features) while preserving them, including identity information, for automatic processing.

Gafni et al. \cite{119} proposed a face de-identification method aimed at maximizing the disassociation of identity while maintaining perceptual invariance (pose, lighting, and expression). Unlike other adversarial perturbations, this approach does not distort the mid-level features of the image. Li et al. \cite{124} introduced a method designed to distort image data in order to hinder classifier inference without affecting the utility for social media users. Sun et al. \cite{126} developed a framework based on the concept of minimal visually significant difference to generate adversarial privacy-preserving images that have minimal perceptual differences from clean images but can attack deep learning models. Jia et al. \cite{127} combined image watermarking technology with adversarial example algorithms to propose a novel watermark perturbation algorithm for adversarial examples. Adding meaningful watermarks to clean images can attack deep neural network models.

Qu et al. \cite{111} proposed a robust adversarial perturbation approach to address the impact of lossy compression on perturbations in online social networks. This scheme design a compression approximation model to learn the ability to disrupt adversarial perturbations, and use this model to guide the generation of robust perturbations, effectively protecting face images under complex OSN compression from Deepfakes, especially for OSNs employing stricter compression. In online social networks, multiple images can not only reflect specific content but also expose the location privacy of sharers. To mitigate the risks associated with location inference, Ma et al. \cite{113} proposed a system that adds defensive perturbations to images to mitigate attacks by creating perturbed images that evade adversarial inference. Unlike blurring or cropping, which makes images less visible to both adversaries and users, this scheme preserves visual content for benign users. To address privacy concerns arising from deep hashing-based facial image retrieval, Tang et al. \cite{112} were the first to propose universally transferable adversarial perturbations. A single adversarial perturbation can be applied to images of all users in the database, while demonstrating transferability to black-box models and unknown hashing schemes. Additionally, the first black-box video attack framework was also proposed \cite{123}.

Physiological signals such as heartbeats and breathing can be remotely captured from human faces using ordinary color cameras under ambient light, which may raise a series of privacy concerns \cite{120}. Therefore, Chen et al. \cite{121} implemented a method that can edit physiological signals in facial videos without affecting the visual appearance, thereby protecting users' physiological signals from leakage and without introducing noticeable visual distortions in the videos. Subsequently, Sun et al. \cite{122} proposed a new method based on a pre-trained 3D convolutional neural network for modifying physiological signals in facial videos to protect privacy.

\begin{table}
\centering

\caption{A Summary of the Significant Change Solutions in the Feature Level}
\scalebox{0.9}{
\begin{tabular}{cccccccccc} 
\toprule
\multirow{2}{*}{Sub-class}         & \multirow{2}{*}{Paper } & \multirow{2}{*}{Year} & \multirow{2}{*}{Key Technique} & \multirow{2}{*}{NI } & \multicolumn{2}{l}{\makecell[c]{PC}} & \multicolumn{2}{l}{\makecell[c]{PC}} & \multirow{2}{*}{Task }  \\ 
\cmidrule{6-7}
\cmidrule{8-9}
                           &                     &                     &                     &                     &                    C &   P                  & H & C                 &                      \\ 
\midrule
\multirow{7}{*}{\makecell[c]{Feature-Level\\ Reversibility}  } & \cite{88}               & 2021                & Synthesis                   &\ding{51}                   & ID                    & NSA        & \ding{51}  & \ding{51}                  & Recognition \\
                           & \cite{89}              & 2021                & Synthesis                    &\ding{51}                   & ID                    & NSA        & \ding{51}  & \ding{51}                  & Recognition \\
                           & \cite{90}              & 2022                & Synthesis                                      &\ding{51}                   & ID                    & NSA        & \ding{51}  & \ding{51}                  & Recognition \\
                           & \cite{91}              & 2023                & Synthesis                                     &\ding{51}                   & ID                    & NSA        & \ding{51}  & \ding{51}                  & Recognition \\
                           & \cite{92}              & 2023                & Synthesis                                      &\ding{51}                   & NSA                    & ID        & \ding{51}  & \ding{51}                  & Recognition \\
                           & \cite{93}              & 2024                & Synthesis                                     &\ding{51}                   & ID                    & NSA        & \ding{51}  & \ding{51}                  & Recognition \\
                           & \cite{104}              & 2022                & Synthesis                                     &\ding{51}                   & ID                    & Expression        & \ding{51}  & \ding{51}                  & Recognition \\
                           
\midrule
\multirow{8}{*}{\makecell{Feature-Level\\ Irreversibility}  }        & \cite{94}              & 2019                &Synthesis   & \ding{53}                    & ID            &Expression                    &\ding{51} &\ding{51}                   &\makecell{Recognition \\and detection}        \\
                           & \cite{96}              & 2020                &Synthesis & \ding{51}                    & ID &Position &\ding{51}   &                  \ding{51}                     &Recognition                    \\
                           & \cite{97}              & 2021                &Synthesis                     &                     \ding{51}                    & ID            &\makecell{Expression\\ Gender}                                      &\ding{51}   &                  \ding{51}      & Recognition  \\
                           & \cite{98}              & 2021                & Synthesis    &                     \ding{51}                    & ID            &NSA                                      &\ding{51}   &                  \ding{51}      & Recognition            \\
                           & \cite{99}              & 2020                &Synthesis                     &                     \ding{53}                    & \makecell{Age, Sex\\ Race}            &NSA                                     &\ding{51}   &                  \ding{51}      & Recognition  \\
                           & \cite{100}              & 2021                &Synthesis                     &                     \ding{51}                    & \makecell{Specifying \\Attributes}           &NSA                                      &\ding{51}   &                  \ding{51}      & Recognition \\
                           & \cite{101}              & 2023                &Synthesis                     &                     \ding{53}                    & ID            &NSA                                      &\ding{51}   &                  \ding{51}      & Recognition \\
                           & \cite{105}              & 2021                &Synthesis                     &                     \ding{51}                    & ID            &NSA                                      &\ding{51}   &                  \ding{51}      & Segmentation \\

\bottomrule
\end{tabular}}

 \begin{tablenotes}
        \footnotesize
        \item[*] NI: naturalness of the image; AC: attribute classification; C: target attributes to be preserved; P: target attributes to be concealed; CT: confrontation type; H: human vision; C: computer vision; NSA: no specific attribute.  
      \end{tablenotes}
\end{table}

Adversarial perturbations generally suffer from poor image quality, limiting the application of these methods in real-world scenarios. Therefore, Lyu et al. \cite{128} presented a 3D-aware adversarial synthetic generation GAN, aimed at improving the quality and portability of identity-concealing synthetic faces. The generation module comprises a novel makeup adjustment module and makeup transfer module, which can leverage the symmetry of human faces to render realistic and robust makeup. Subsequently, Shamshad et al. \cite{130} introduced a novel two-step method to search for adversarial latent codes that can be utilized by generative models (such as StyleGAN) to produce face images with high visual quality that match human-perceived identities while simultaneously deceiving black-box face recognition systems.

\subsection{Design Principles}
Tables 5 and 6 respectively provide a detailed classification of solutions for feature-level privacy protection, distinguishing between those that cause significant changes to images and those that cause non-significant changes. Based on the detailed descriptions and classifications of various schemes mentioned above, we summarize several common design principles for this type of privacy protection methods.

%
%

\begin{table}
\centering
\caption{A Summary of the Non-Significant Change Solutions in the Feature Level}
\scalebox{0.9}{
\begin{tabular}{cccccccccc} 
\toprule
\multirow{2}{*}{Sub-class}         & \multirow{2}{*}{Paper } & \multirow{2}{*}{Year} & \multirow{2}{*}{Key Technique} & \multirow{2}{*}{NI } & \multicolumn{2}{l}{\makecell[c]{PC}} & \multicolumn{2}{l}{\makecell[c]{PC}} & \multirow{2}{*}{Task }  \\ 
\cmidrule{6-7}
\cmidrule{8-9}
                           &                     &                     &                     &                     &                    C &   P                  & H & C                 &                      \\ 
\midrule
\multirow{4}{*}{\makecell{Feature-Level\\ Reversibility}  } & \cite{102}               & 2024                & Perturbation                   &\ding{51}                   & Sex                    & Other        & \ding{53}  & \ding{51}                  & Recognition \\
                           & \cite{106}              & 2024                & Perturbation                    &\ding{51}                   & N/A                    &N/A         & \ding{53}  & \ding{51}                  & Copyright \\
                           & \cite{107}              & 2023                & Perturbation                                      &\ding{53}                   & ID                    & Other        & \ding{53}  & \ding{51}                  & Recognition \\
                           & \cite{108}              & 2023                & Perturbation                                     &\ding{51}                   & N/A                    & N/A        & \ding{53}  & \ding{51}                  & Recognition \\
                           
\midrule
\multirow{16}{*}{\makecell{Feature-Level\\ Irreversibility}  }        & \cite{109}              & 2021                &Perturbation   & \ding{53}                    & ID            &Other                    &\ding{53} &\ding{51}                   &Recognition        \\
                           & \cite{110}              & 2022                & Perturbation                    &\ding{53}                   & ID                    &Other         & \ding{53}  & \ding{51}                  & Recognition \\
                           & \cite{125}              & 2024                & Perturbation                                      &\ding{53}                   & ID                    & Other        & \ding{53}  & \ding{51}                  & Recognition \\
                           & \cite{114}              & 2023                & Perturbation                                     &\ding{53}                   & Text,ID                    & Other        & \ding{53}  & \ding{51}                  & Recognition \\
                           & \cite{115}              & 2021                & Perturbation                    &\ding{51}                   & N/A                    &N/A         & \ding{53}  & \ding{51}                  & Detection \\
                           & \cite{117}              & 2021                & Perturbation                                     &\ding{51}                   & ID                    & Other        & \ding{53}  & \ding{51}                  & Recognition \\
                           & \cite{126}              & 2023                & Perturbation                    &\ding{51}                   & N/A                    &N/A         & \ding{53}  & \ding{51}                  & \makecell{Recognition,\\Classification} \\
                           & \cite{127}              & 2020                & Perturbation                                      &\ding{51}                   & N/A                    & N/A        & \ding{53}  & \ding{51}                  & Recognition \\
                           & \cite{130}              & 2023                & Perturbation                                     &\ding{51}                   & ID                    & Other        & \ding{53}  & \ding{51}                  & Recognition \\
                           & \cite{111}              & 2024                & Perturbation                    &\ding{51}                   & ID                    &Other         & \ding{53}  & \ding{51}                  & Recognition \\
                           & \cite{113}              & 2024                & Perturbation                                      &\ding{51}                   & \makecell{Geographic\\ location}                & Other        & \ding{53}  & \ding{51}                  & Inference \\
                           & \cite{112}              & 2024                & Perturbation                                     &\ding{53}                   & ID                    & Other        & \ding{53}  & \ding{51}                  & Retrieval \\
                           & \cite{121}              & 2022                & Perturbation                                      &\ding{51}                   & \makecell{Physiological\\ signals}     & Other        & \ding{53}  & \ding{51}                  & Detection \\
                           & \cite{122}              & 2022                & Perturbation                                     &\ding{51}                   & \makecell{Physiological\\ signals}   & Other        & \ding{53}  & \ding{51}                  & Detection \\
                           & \cite{128}              & 2023                & Perturbation                    &\ding{51}                   & ID                    &Other         & \ding{53}  & \ding{51}                  & Recognition \\

\bottomrule
\end{tabular}}

 \begin{tablenotes}
        \footnotesize
        \item[*] NI: naturalness of the image; AC: attribute classification; C: target attributes to be preserved; P: target attributes to be concealed; CT: confrontation type; H: human vision; C: computer vision; NSA: no specific attribute.  
      \end{tablenotes}
\end{table}

\begin{itemize}
\item {\bfseries{High visual usability.}} The core of feature-level privacy protection lies in fine-tuning specific parts of the image that may reveal sensitive information at the semantic level, rather than making extensive and crude modifications to the entire image content. This allows observers to browse the image without noticing any unnaturalness, blurring, deletion traces, or mismatched replacements. Therefore, it can effectively eliminate potential privacy leakage risks while maximizing the retention of image authenticity, accuracy, and visual usability.

\item {\bfseries{Adversarial machines.}} Before tackling the content of image privacy, a crucial preliminary issue is to identify the potential sources of threat, i.e., to determine who the “adversaries” are. In the complex environment of image transmission, threats to privacy invasion may stem from human curiosity and misconduct, or may lurk within highly intelligent artificial intelligence systems, or even both may coexist. For data-level and content-level privacy protection strategies, the traditional focus of protection often centers on resisting the risk of direct human recognition to ensure that sensitive information is not easily perceived. However, in the context of feature-level privacy protection, we must consider machines as the most critical challenge.

\item {\bfseries{Machine privacy.}} In image privacy protection, beyond the intuitive and easily recognizable traditional privacy content, there is a more advanced, subtle privacy dimension that can only be interpreted by machine intelligence—machine privacy. This type of privacy information transcends the limitations of visual observation and conventional prior knowledge. Its existence is often not easily detected \cite{138} but can be fully exposed through complex algorithms and model analysis. When exploring strategies for feature-level privacy protection, we particularly focus on protecting this level of machine privacy.

\end{itemize}

\section{CHALLENGES AND FUTURE DIRECTION}
\subsection{Dynamic Revocable Image Privacy Protection}
In reviewing current privacy protection technologies, a core premise lies in the clear definition of privacy objectives, a concept inherently subjective and context-dependent. For individuals sensitive to geographical locations, information such as landmark buildings, regional activity indicators, and street layouts depicted in images are often considered primary targets for privacy protection. In contrast, for those sensitive to facial features, facial data in images constitute the main privacy concern. However, existing strategies often rely on pre-set privacy objectives by researchers or developers for protection design. While such methods can effectively safeguard against specific privacy needs, they struggle to generalize to broader privacy contexts, particularly neglecting individual differences among users and their dynamically changing privacy preferences.

This limitation highlights the urgency of designing user-customizable privacy protection mechanisms. Ideally, a flexible framework should be developed that allows users to dynamically define and evaluate what is private information and what is publicly releasable based on different image contents, specific usage scenarios, and even personal preferences. The protection of the old definition of privacy needs to be withdrawn until the new definition of privacy is protected. The framework needs to integrate a set of quantitative indicators or an assessment system to scientifically and systematically define privacy boundaries in order to ensure the relevance and effectiveness of privacy protection measures.

\subsection{User-Understandable Image Privacy Protection}
User-understandable privacy preservation aims at targeting the balance between privacy and usability. Although we have analyzed the privacy and usability of images in our previous review analysis, the issue of balancing usability and privacy is different for different levels of privacy preservation, which we will explore further here. The essence of an image lies in recording information for oneself or visually conveying what words cannot express to others, which can also be regarded as its utility. While privacy preserving methods enhance the privacy of images, they concurrently impair this utility, sometimes even resulting in its complete loss. Thus, to what extent is the visual content of an image preserved in order to be considered as having better utility and privacy? Until now, few indicators have been able to explain the relationship between the two. The latest review \cite{134} addressed these points but still did not provide a particularly clear answer. This unclear balance can have a significant impact on daily life. If we want to share photos, assuming we take privacy to the extreme and leave no visual usability, the original purpose of the initial transmission of the image goes up in smoke. Would we still have the desire to share images on social platforms? The answer is definitely no. Therefore, the balance between privacy and usability is one of our main focuses as well as one of our most difficult challenges.

As a carrier of visual information, the core value of images lies in conveying the richness of details and emotions that are difficult to capture precisely in words, which constitutes the inherent utility of images. However, while the implementation of privacy protection strategies aims to enhance the security of image data, it will inevitably erode this utility to a certain extent, and in extreme cases, may even completely sacrifice its function of information transmission.

Currently, academics still face many challenges in exploring the optimal balance between image privacy and utility, and the ambiguity of this balance relationship has far-reaching impacts on daily life and social media applications. In image sharing scenarios, if privacy protection is overemphasized and the visual usability of images is completely deprived, it violates the basic purpose of image sharing, which is to exchange information and share emotions, and thus weakens users' willingness and motivation to share images. Therefore, how to maximize the utility of images while protecting personal privacy has become an important challenge in the field of image processing and privacy protection.

\subsection{Image Privacy Enhancement under Multimodal Learning}
In the context of multimodal learning, image privacy enhancement becomes particularly important. Specifically, through multimodal learning, the system can infer private information in an image from relatively innocuous modalities such as text or speech. For example, a particular image may contain only partially recognizable content. However, when combined with text descriptions or voice prompts, the system may be able to accurately recognize the personal information implied behind the image, such as identity, location, activity, etc. Compared with a single modality (e.g., image), multimodal data (including text, speech, and video, etc.) carries different types of information, the combination of which can lead to richer context and situation understanding.

Importantly, this fusion of multimodal data also makes privacy protection a more complex challenge. For example, images alone may not be able to tell if a person is in a sensitive place or has had contact with certain people. However, when images are fused with other modal data, such as text and voice, the information can complement each other to form a more complete and detailed portrait of a person's personal information, exposing privacy that would otherwise be undetectable. This increased privacy risk requires us to pay extra attention to the innovation and strengthening of privacy protection mechanisms in the design and application of multimodal learning. Therefore, how to effectively protect user privacy while enjoying the convenience of multimodal learning has become a major challenge and research focus in privacy protection.

\section{CONCLUSION}
The rapid growth and wide dissemination of images have been extensively used in various fields. However, images may face the risk of being copied, tampered with and improperly distributed during storage and transmission. This makes the private information implicit in images vulnerable to leakage, which in turn poses potential security risks. This survey comprehensively discusses related research in the area of image privacy protection, and reviews and analyzes existing solutions. Specifically, we categorize the privacy content of images and propose the concept of “privacy-sensitive domain” to describe the privacy-related regions in images. Based on the privacy-sensitive domain, we construct a comprehensive image privacy protection framework that covers multiple application scenarios and privacy protection goals. Then, we categorize existing image privacy protection methods into data-level, content-level, and feature-level privacy protection, and discuss in detail the main methods and technical features in each category. Finally, this survey looks at the possible directions of development in the field of image privacy protection and explores future research challenges and opportunities.

\bibliographystyle{unsrt}
\bibliography{acmart}

\begin{thebibliography}{100}

\bibitem{139}
Aristotle.
\newblock {\em Metaphysics}, volume~1.
\newblock 1933.

\bibitem{140}
Dhiraj Joshi, Ritendra Datta, Elena Fedorovskaya, Quang-Tuan Luong, James~Z
  Wang, Jia Li, and Jiebo Luo.
\newblock Aesthetics and emotions in images.
\newblock {\em IEEE Signal Processing Magazine}, 28(5):94--115, 2011.

\bibitem{141}
Martha~L King and Victor~M Rentel.
\newblock Research update: Conveying meaning in written texts.
\newblock {\em Language Arts}, 58(6):721--728, 1981.

\bibitem{142}
Jason Lankow, Josh Ritchie, and Ross Crooks.
\newblock {\em Infographics: The power of visual storytelling}.
\newblock John Wiley \& Sons, 2012.

\bibitem{143}
John Berger.
\newblock {\em Ways of seeing}.
\newblock Penguin uK, 2008.

\bibitem{144}
Tess Van~der Zanden, Maria~BJ Mos, Alexander~P Schouten, and Emiel~J Krahmer.
\newblock What people look at in multimodal online dating profiles: How
  pictorial and textual cues affect impression formation.
\newblock {\em Communication Research}, 49(6):863--890, 2022.

\bibitem{145}
Suhad~Ibraheem Kadhem, Intisar~AM Al~Sayed, Thuria~Saad Znad, Jamal~Fadhil
  Tawfeq, Ahmed~Dheyaa Radhi, Hassan~Muwafaq Gheni, and Israa Al-Barazanchi.
\newblock A review of {WBAN} intelligent system connections for remote control
  of patients with {COVID}-19.
\newblock In {\em The International Conference of Advanced Computing and
  Informatics}, pages 242--254, 2022.

\bibitem{2}
Chi Liu, Tianqing Zhu, Jun Zhang, and Wanlei Zhou.
\newblock Privacy intelligence: A survey on image privacy in online social
  networks.
\newblock {\em ACM Computing Surveys (CSUR)}, 55(8):1--35, 2022.

\bibitem{132}
M~Shamim Hossain and Ghulam Muhammad.
\newblock Cloud-assisted speech and face recognition framework for health
  monitoring.
\newblock {\em Mobile Networks and Applications}, 20:391--399, 2015.

\bibitem{133}
Yilun Wang and Michal Kosinski.
\newblock Deep neural networks are more accurate than humans at detecting
  sexual orientation from facial images.
\newblock {\em Journal of Personality and Social Psychology}, 114(2):246, 2018.

\bibitem{3}
Jos{\'e} Alemany, E~Del Val, and Ana Garc{\'\i}a-Fornes.
\newblock A review of privacy decision-making mechanisms in online social
  networks.
\newblock {\em ACM Computing Surveys (CSUR)}, 55(2):1--32, 2022.

\bibitem{4}
Jos{\'e}~Ram{\'o}n Padilla-L{\'o}pez, Alexandros~Andre Chaaraoui, and Francisco
  Fl{\'o}rez-Revuelta.
\newblock Visual privacy protection methods: A survey.
\newblock {\em Expert Systems with Applications}, 42(9):4177--4195, 2015.

\bibitem{7}
Thomas Winkler and Bernhard Rinner.
\newblock Security and privacy protection in visual sensor networks: A survey.
\newblock {\em ACM Computing Surveys (CSUR)}, 47(1):1--42, 2014.

\bibitem{5}
Md~Rezwan Hasan, Richard Guest, and Farzin Deravi.
\newblock Presentation-level privacy protection techniques for automated face
  recognition—a survey.
\newblock {\em ACM Computing Surveys (CSUR)}, 55(13s):1--27, 2023.

\bibitem{6}
Bla{\v{z}} Meden, Peter Rot, Philipp Terh{\"o}rst, Naser Damer, Arjan Kuijper,
  Walter~J Scheirer, Arun Ross, Peter Peer, and Vitomir {\v{S}}truc.
\newblock Privacy--enhancing face biometrics: A comprehensive survey.
\newblock {\em IEEE Transactions on Information Forensics and Security},
  16:4147--4183, 2021.

\bibitem{8}
Georgios~A Kaissis, Marcus~R Makowski, Daniel R{\"u}ckert, and Rickmer~F
  Braren.
\newblock Secure, privacy-preserving and federated machine learning in medical
  imaging.
\newblock {\em Nature Machine Intelligence}, 2(6):305--311, 2020.

\bibitem{10}
Zezong Zhang, Jianeng Tang, Feng Zhang, Tingting Huang, and Mingsheng Lu.
\newblock Medical image encryption based on josephus scrambling and dynamic
  cross-diffusion for patient privacy security.
\newblock {\em IEEE Transactions on Circuits and Systems for Video Technology},
  34(10):9250--9263, 2024.

\bibitem{11}
Satendra~Pal Singh and Gaurav Bhatnagar.
\newblock A novel biometric inspired robust security framework for medical
  images.
\newblock {\em IEEE Transactions on Knowledge and Data Engineering},
  33(3):810--823, 2019.

\bibitem{9}
Yanli Chen, Tong Qiao, Florent Retraint, and Gengran Hu.
\newblock Efficient privacy-preserving forensic method for camera model
  identification.
\newblock {\em IEEE Transactions on Information Forensics and Security},
  17:2378--2393, 2022.

\bibitem{12}
Pauline Puteaux and William Puech.
\newblock Cfb-then-ecb mode-based image encryption for an efficient correction
  of noisy encrypted images.
\newblock {\em IEEE Transactions on Circuits and Systems for Video Technology},
  31(9):3338--3351, 2020.

\bibitem{13}
Yongjin Xian, Xingyuan Wang, and Lin Teng.
\newblock Double parameters fractal sorting matrix and its application in image
  encryption.
\newblock {\em IEEE Transactions on Circuits and Systems for Video Technology},
  32(6):4028--4037, 2021.

\bibitem{14}
MA~S{\'a}nchez-Granero and M~Fern{\'a}ndez-Mart{\'\i}nez.
\newblock Irreducible fractal structures for moran type theorems.
\newblock {\em Chaos, Solitons \& Fractals}, 119:29--36, 2019.

\bibitem{17}
Sicong Liu, Chunbiao Li, and Yongxing LI.
\newblock A novel image encryption algorithm based on exponent-cosine chaotic
  mapping.
\newblock {\em Journal of Electronics Information Technology},
  44(5):1754--1762, 2022.

\bibitem{15}
Xingyuan Wang and Suo Gao.
\newblock Image encryption algorithm based on the matrix semi-tensor product
  with a compound secret key produced by a boolean network.
\newblock {\em Information Sciences}, 539:195--214, 2020.

\bibitem{16}
Yuwen Sha, Jun Mou, Santo Banerjee, and Yushu Zhang.
\newblock Exploiting flexible and secure cryptographic technique for
  multi-dimensional image based on graph data structure and three-input
  majority gate.
\newblock {\em IEEE Transactions on Industrial Informatics}, 20(3):3835--3846,
  2024.

\bibitem{29}
Yuan Yuan, Hongjie He, Yaolin Yang, Hadi Amirpour, Christian Timmerer, and Fan
  Chen.
\newblock Jpeg image encryption with {DC} rotation and undivided {RSV}-based
  {AC} group permutation.
\newblock {\em IEEE Transactions on Multimedia}, 2023.
\newblock doi: 10.1109/TMM.2023.3336236.

\bibitem{18}
Bo~Zhang, Di~Xiao, and Yong Xiang.
\newblock Robust coding of encrypted images via 2{D} compressed sensing.
\newblock {\em IEEE Transactions on Multimedia}, 23:2656--2671, 2020.

\bibitem{20}
Jia Liang, Di~Xiao, Hui Huang, and Min Li.
\newblock Multilevel privacy preservation scheme based on compressed sensing.
\newblock {\em IEEE Transactions on Industrial Informatics}, 19(6):7435--7444,
  2022.

\bibitem{21}
Xiuli Chai, Jiangyu Fu, Zhihua Gan, Yang Lu, Yushu Zhang, and Daojun Han.
\newblock Exploiting semi-tensor product compressed sensing and hybrid cloud
  for secure medical image transmission.
\newblock {\em IEEE Internet of Things Journal}, 10(8):7380--7392, 2022.

\bibitem{22}
lan Wang, Di~Xiao, Fei Wang, and Xi~Shi.
\newblock Image robust encryption algorithm based on scrambled block
  compressive sensing.
\newblock {\em Journal of Cryptologic Research}, 9(2):267--283, 2022.

\bibitem{23}
Mengxin Gong, Xiuli Chai, Yang Lu, and Yushu Zhang.
\newblock Exploiting four-dimensional chaotic systems with dissipation and
  optimized logical operations for secure image compression and encryption.
\newblock {\em IEEE Transactions on Circuits and Systems for Video Technology},
  34(8):7628--7642, 2024.

\bibitem{24}
Xinyu Gao, Jun Mou, Santo Banerjee, and Yushu Zhang.
\newblock Color-gray multi-image hybrid compression-encryption scheme based on
  {BP} neural network and knight tour.
\newblock {\em IEEE Transactions on Cybernetics}, 53(8):5037--5047, 2023.

\bibitem{25}
Chuntao Wang, Juan Hu, Shan Bian, Jiangqun Ni, and Xinpeng Zhang.
\newblock A customized deep network based encryption-then-lossy-compression
  scheme of color images achieving arbitrary compression ratios.
\newblock {\em IEEE Transactions on Circuits and Systems for Video Technology},
  33(8):4322--4336, 2023.

\bibitem{26}
Feng Wang, Zhuhong Shao, Yunfei Wang, Qijun Yao, and Xilin Liu.
\newblock Multiple image encryption of high robustness in gyrator transform
  domain.
\newblock {\em Journal of Image and Graphics}, 25(7):1366--1379, 2020.

\bibitem{27}
Chuan Qin, Jinchuan Hu, Fengyong Li, Zhenxing Qian, and Xinpeng Zhang.
\newblock Jpeg image encryption with adaptive {DC} coefficient prediction and
  {RS} pair permutation.
\newblock {\em IEEE Transactions on Multimedia}, 25:2528--2542, 2022.

\bibitem{28}
Qihua Feng, Peiya Li, Zhixun Lu, Chaozhuo Li, Zefang Wang, Zhiquan Liu, Chunhui
  Duan, Feiran Huang, Jian Weng, et~al.
\newblock Evit: Privacy-preserving image retrieval via encrypted vision
  transformer in cloud computing.
\newblock {\em IEEE Transactions on Circuits and Systems for Video Technology},
  34(8):7467--7483, 2024.

\bibitem{30}
Dan Zhu, Hui Zhu, Xiangyu Wang, Rongxing Lu, and Dengguo Feng.
\newblock An accurate and privacy-preserving retrieval scheme over outsourced
  medical images.
\newblock {\em IEEE Transactions on Services Computing}, 16(2):913--926, 2022.

\bibitem{31}
J~Anju and R~Shreelekshmi.
\newblock A secure image outsourcing using privacy-preserved local color layout
  descriptor in cloud environment.
\newblock {\em IEEE Transactions on Services Computing}, 17(2):378--391, 2024.

\bibitem{32}
Dong Li, Qingguo L{\"u}, Xiaofeng Liao, Tao Xiang, Jiahui Wu, and Junqing Le.
\newblock Avpmir: Adaptive verifiable privacy-preserving medical image
  retrieval.
\newblock {\em IEEE Transactions on Dependable and Secure Computing},
  21(5):4637--4651, 2024.

\bibitem{135}
Zixuan Yang, Yushu Zhang, Tao Wang, Zhongyun Hua, Zhihua Xia, and Jian Weng.
\newblock Once-for-all: Efficient visual face privacy protection via
  person-specific veils.
\newblock In {\em Proceedings of the 32nd ACM International Conference on
  Multimedia}, pages 7705--7713, 2024.

\bibitem{33}
Yi~Ding, Guozheng Wu, Dajiang Chen, Ning Zhang, Linpeng Gong, Mingsheng Cao,
  and Zhiguang Qin.
\newblock {DeepEDN}: A deep-learning-based image encryption and decryption
  network for internet of medical things.
\newblock {\em IEEE Internet of Things Journal}, 8(3):1504--1518, 2020.

\bibitem{34}
Enjun Zhu, Haiyu Feng, Long Chen, Yongqiang Lai, and Senchun Chai.
\newblock {MP-Net}: A multi-center privacy-preserving network for medical image
  segmentation.
\newblock {\em IEEE Transactions on Medical Imaging}, 43(7):2718--2729, 2024.

\bibitem{35}
Hao Wu, Xuejin Tian, Yuhang Gong, Xing Su, Minghao Li, and Fengyuan Xu.
\newblock {DAPter}: Preventing user data abuse in deep learning inference
  services.
\newblock In {\em Proceedings of the Web Conference 2021}, pages 1017--1028,
  2021.
\newblock doi: {10.1145/3442381.3449907}.

\bibitem{36}
Jun Liu, Jiantao Zhou, Jinyu Tian, and Weiwei Sun.
\newblock Recoverable privacy-preserving image classification through
  noise-like adversarial examples.
\newblock {\em ACM Transactions on Multimedia Computing, Communications and
  Applications}, 20(7):1--27, 2024.

\bibitem{37}
Zhigang Su, Dawei Zhou, Nannan Wang, Decheng Liu, Zhen Wang, and Xinbo Gao.
\newblock Hiding visual information via obfuscating adversarial perturbations.
\newblock In {\em Proceedings of the IEEE/CVF International Conference on
  Computer Vision}, pages 4356--4366, 2023.

\bibitem{38}
Alem Fitwi, Yu~Chen, and Sencun Zhu.
\newblock Enforcing privacy preservation on edge cameras using lightweight
  video frame scrambling.
\newblock {\em IEEE Transactions on Services Computing}, 16(1):276--287, 2021.

\bibitem{39}
Liyun Dou, Zichi Wang, Zhenxing Qian, and Guorui Feng.
\newblock Reversible privacy protection with the capability of antiforensics.
\newblock {\em Security and Communication Networks}, 2021(1), 2021.

\bibitem{40}
Pengbo Liu, Xingyuan Wang, and Yining Su.
\newblock Image encryption via complementary embedding algorithm and new
  spatiotemporal chaotic system.
\newblock {\em IEEE Transactions on Circuits and Systems for Video Technology},
  33(5):2506--2519, 2022.

\bibitem{41}
Terrance~Edward Boult.
\newblock Pico: Privacy through invertible cryptographic obscuration.
\newblock In {\em Computer Vision for Interactive and Intelligent Environment
  (CVIIE'05)}, pages 27--38, 2005.

\bibitem{45}
Xin Cao, Yuxuan Huang, Hao-Tian Wu, and Yiu-ming Cheung.
\newblock Content and privacy protection in {JPEG} images by reversible visual
  transformation.
\newblock {\em Applied Sciences}, 10(19):6776, 2020.

\bibitem{48}
Natacha Ruchaud and Jean-Luc Dugelay.
\newblock Aseppi: Robust privacy protection against {De-Anonymization} attacks.
\newblock In {\em 2017 IEEE Conference on Computer Vision and Pattern
  Recognition Workshops (CVPRW)}, pages 1352--1359, 2017.
\newblock doi: {10.1109/CVPRW.2017.177}.

\bibitem{19}
Xiaofei He, Lixiang Li, Fenghua Tong, and Haipeng Peng.
\newblock Multi-level privacy protection for social media based on 2{D}
  compressive sensing.
\newblock {\em IEEE Internet of Things Journal}, 11(4):6878--6892, 2024.

\bibitem{42}
Lin Yuan and Touradj Ebrahimi.
\newblock Image privacy protection with secure {JPEG} transmorphing.
\newblock {\em IET Signal Processing}, 11(9):1031--1038, 2017.

\bibitem{46}
Lin Yuan and Touradj Ebrahimi.
\newblock Image transmorphing with {JPEG}.
\newblock In {\em 2015 IEEE International Conference on Image Processing
  (ICIP)}, pages 3956--3960, 2015.

\bibitem{43}
Lin Yuan, Pavel Korshunov, and Touradj Ebrahimi.
\newblock Privacy-preserving photo sharing based on a secure {JPEG}.
\newblock In {\em 2015 IEEE Conference on Computer Communications Workshops
  (INFOCOM WKSHPS)}, pages 185--190, 2015.

\bibitem{44}
Zhengxin You, Sheng Li, Zhenxing Qian, and Xinpeng Zhang.
\newblock Reversible privacy-preserving recognition.
\newblock In {\em 2021 IEEE International Conference on Multimedia and Expo
  (ICME)}, pages 1--6, 2021.

\bibitem{49}
Jun Yu, Baopeng Zhang, Zhengzhong Kuang, Dan Lin, and Jianping Fan.
\newblock iprivacy: Image privacy protection by identifying sensitive objects
  via deep multi-task learning.
\newblock {\em IEEE Transactions on Information Forensics and Security},
  12(5):1005--1016, 2016.

\bibitem{52}
Yuanyi Sun, Sencun Zhu, and Yu~Chen.
\newblock {ZoomP3}: Privacy-preserving publishing of online video conference
  recordings.
\newblock {\em Proceedings on Privacy Enhancing Technologies}, 2022:630--649,
  2022.

\bibitem{53}
Jizhe Zhou and Chi-Man Pun.
\newblock Personal privacy protection via irrelevant faces tracking and
  pixelation in video live streaming.
\newblock {\em IEEE Transactions on Information Forensics and Security},
  16:1088--1103, 2020.

\bibitem{129}
Tribhuvanesh Orekondy, Mario Fritz, and Bernt Schiele.
\newblock Connecting pixels to privacy and utility: Automatic redaction of
  private information in images.
\newblock In {\em Proceedings of the IEEE Conference on Computer Vision and
  Pattern Recognition (CVPR)}, June 2018.

\bibitem{51}
Baowei Jiang, Bing Bai, Haozhe Lin, Yu~Wang, Yuchen Guo, and Lu~Fang.
\newblock Dartblur: Privacy preservation with detection artifact suppression.
\newblock In {\em Proceedings of the IEEE/CVF Conference on Computer Vision and
  Pattern Recognition}, pages 16479--16488, 2023.

\bibitem{50}
Nadiya Shvai, Arcadi~Llanza Carmona, and Amir Nakib.
\newblock Adaptive image anonymization in the context of image classification
  with neural networks.
\newblock In {\em Proceedings of the IEEE/CVF International Conference on
  Computer Vision}, pages 5074--5083, 2023.

\bibitem{54}
Joshua Morris, Sara Newman, Kannappan Palaniappan, Jianping Fan, and Dan Lin.
\newblock “do you know you are tracked by photos that you didn't take”:
  Large-scale location-aware multi-party image privacy protection.
\newblock {\em IEEE Transactions on Dependable and Secure Computing},
  20(1):301--312, 2021.

\bibitem{55}
Yahan Yang, Junfeng Lyu, Ruixin Wang, Quan Wen, Lanqin Zhao, Wenben Chen,
  Shaowei Bi, Jie Meng, Keli Mao, Yu~Xiao, et~al.
\newblock A digital mask to safeguard patient privacy.
\newblock {\em Nature Medicine}, 28(9):1883--1892, 2022.

\bibitem{56}
Karla Brkic, Ivan Sikiric, Tomislav Hrkac, and Zoran Kalafatic.
\newblock I know that person: Generative full body and face de-identification
  of people in images.
\newblock In {\em 2017 IEEE Conference on Computer Vision and Pattern
  Recognition Workshops (CVPRW)}, pages 1319--1328, 2017.

\bibitem{57}
Charles~V Wright, Wu-chi Feng, and Feng Liu.
\newblock Thumbnail-preserving encryption for {JPEG}.
\newblock In {\em Proceedings of the 3rd ACM Workshop on Information Hiding and
  Multimedia Security}, pages 141--146, 2015.

\bibitem{58}
Kimia Tajik, Akshith Gunasekaran, Rhea Dutta, Brandon Ellis, Rakesh~B Bobba,
  Mike Rosulek, Charles~V Wright, and Wu-chi Feng.
\newblock Balancing image privacy and usability with thumbnail-preserving
  encryption.
\newblock In {\em Network and Distributed System Security Symposium}.
\newblock doi: {10.14722/ndss.2019.23432}.

\bibitem{59}
Ruoyu Zhao, Yushu Zhang, Xiangli Xiao, Xi~Ye, and Rushi Lan.
\newblock Tpe2: Three-pixel exact thumbnail-preserving image encryption.
\newblock {\em Signal Processing}, 183:108019, 2021.

\bibitem{66}
Ruoyu Zhao, Yushu Zhang, Rushi Lan, Zhongyun Hua, and Yong Xiang.
\newblock Heterogeneous and customized cost-efficient reversible image
  degradation for green {IoT}.
\newblock {\em IEEE Internet of Things Journal}, 10(3):2630--2645, 2022.

\bibitem{67}
Ruoyu Zhao, Yushu Zhang, Yu~Nan, Wenying Wen, Xiuli Chai, and Rushi Lan.
\newblock Primitively visually meaningful image encryption: A new paradigm.
\newblock {\em Information Sciences}, 613:628--648, 2022.

\bibitem{74}
Yushu Zhang, Xi~Ye, Xiangli Xiao, Tao Xiang, Hongwei Li, and Xiaochun Cao.
\newblock A reversible framework for efficient and secure visual privacy
  protection.
\newblock {\em IEEE Transactions on Information Forensics and Security},
  18:3334--3349, 2023.

\bibitem{60}
Xi~Ye, Yushu Zhang, Xiangli Xiao, Shuang Yi, and Rushi Lan.
\newblock Usability enhanced thumbnail-preserving encryption based on data
  hiding for {JPEG} images.
\newblock {\em IEEE Signal Processing Letters}, 30:793--797, 2023.

\bibitem{63}
Xiuli Chai, Yakun Ma, Yinjing Wang, Zhihua Gan, and Yushu Zhang.
\newblock {TPE-ADE}: Thumbnail-preserving encryption based on adaptive
  deviation embedding for {JPEG} images.
\newblock {\em IEEE Transactions on Multimedia}, 26:6102--6116, 2024.

\bibitem{64}
Yakun Ma, Xiuli Chai, Zhihua Gan, and Yushu Zhang.
\newblock Privacy-preserving {TPE}-based {JPEG} image retrieval in
  cloud-assisted internet of things.
\newblock {\em IEEE Internet of Things Journal}, 11(3):4842--4856, 2024.

\bibitem{61}
Yushu Zhang, Ruoyu Zhao, Xiangli Xiao, Rushi Lan, Zhe Liu, and Xinpeng Zhang.
\newblock {HF-TPE}: High-fidelity thumbnail-preserving encryption.
\newblock {\em IEEE Transactions on Circuits and Systems for Video Technology},
  32(3):947--961, 2021.

\bibitem{62}
Yushu Zhang, Wentao Zhou, Ruoyu Zhao, Xinpeng Zhang, and Xiaochun Cao.
\newblock {F-TPE}: Flexible thumbnail-preserving encryption based on
  multi-pixel sum-preserving encryption.
\newblock {\em IEEE Transactions on Multimedia}, 25:5877--5891, 2022.

\bibitem{65}
Wenying Wen, Qiyu Jiang, Haigang Huang, Yushu Zhang, and Yuming Fang.
\newblock {TPE-DF}: Thumbnail preserving encryption via dual-{2DCS} fusion.
\newblock {\em IEEE Signal Processing Letters}, 31:1039--1043, 2024.

\bibitem{103}
Mang Ye, Wei Shen, Junwu Zhang, Yao Yang, and Bo~Du.
\newblock Securereid: Privacy-preserving anonymization for person
  re-identification.
\newblock {\em IEEE Transactions on Information Forensics and Security},
  19:2840--2853, 2024.

\bibitem{68}
Shao-Ping Lu, Rong Wang, Tao Zhong, and Paul~L Rosin.
\newblock Large-capacity image steganography based on invertible neural
  networks.
\newblock In {\em Proceedings of the IEEE/CVF Conference on Computer Vision and
  Pattern Recognition}, pages 10816--10825, 2021.

\bibitem{69}
Qichao Ying, Hang Zhou, Xianhan Zeng, Haisheng Xu, Zhenxing Qian, and Xinpeng
  Zhang.
\newblock Hiding images into images with real-world robustness.
\newblock In {\em 2022 IEEE International Conference on Image Processing
  (ICIP)}, pages 111--115, 2022.

\bibitem{70}
Zhenyu Guan, Junpeng Jing, Xin Deng, Mai Xu, Lai Jiang, Zhou Zhang, and Yipeng
  Li.
\newblock {DeepMIH}: Deep invertible network for multiple image hiding.
\newblock {\em IEEE Transactions on Pattern Analysis and Machine Intelligence},
  45(1):372--390, 2022.

\bibitem{71}
Junxue Yang and Xin Liao.
\newblock Exploiting fine-grained {DCT} representations for hiding image-level
  messages within {JPEG} images.
\newblock In {\em Proceedings of the 31st ACM International Conference on
  Multimedia}, pages 7373--7382, 2023.

\bibitem{73}
Guobiao Li, Sheng Li, Zicong Luo, Zhenxing Qian, and Xinpeng Zhang.
\newblock Purified and unified steganographic network.
\newblock In {\em Proceedings of the IEEE/CVF Conference on Computer Vision and
  Pattern Recognition}, pages 27569--27578, 2024.

\bibitem{72}
Jun Jia, Zhongpai Gao, Dandan Zhu, Xiongkuo Min, Guangtao Zhai, and Xiaokang
  Yang.
\newblock Learning invisible markers for hidden codes in offline-to-online
  photography.
\newblock In {\em Proceedings of the IEEE/CVF Conference on Computer Vision and
  Pattern Recognition}, pages 2273--2282, 2022.

\bibitem{75}
Lin Yuan, Kai Liang, Xiao Pu, Yan Zhang, Jiaxu Leng, Tao Wu, Nannan Wang, and
  Xinbo Gao.
\newblock Invertible image obfuscation for facial privacy protection via secure
  flow.
\newblock {\em IEEE Transactions on Circuits and Systems for Video Technology},
  34(7):6077--6091, 2024.

\bibitem{76}
Hao Wu, Xuejin Tian, Minghao Li, Yunxin Liu, Ganesh Ananthanarayanan, Fengyuan
  Xu, and Sheng Zhong.
\newblock {PECAM}: Privacy-enhanced video streaming and analytics via
  securely-reversible transformation.
\newblock In {\em Proceedings of the 27th Annual International Conference on
  Mobile Computing and Networking}, pages 229--241, 2021.

\bibitem{77}
Jun-Yan Zhu, Taesung Park, Phillip Isola, and Alexei~A Efros.
\newblock Unpaired image-to-image translation using cycle-consistent
  adversarial networks.
\newblock In {\em Proceedings of the IEEE International Conference on Computer
  Vision}, pages 2223--2232, 2017.

\bibitem{78}
Serdar {\c{C}}ift{\c{c}}i, Ahmet~O{\u{g}}uz Aky{\"u}z, and Touradj Ebrahimi.
\newblock A reliable and reversible image privacy protection based on false
  colors.
\newblock {\em IEEE Transactions on Multimedia}, 20(1):68--81, 2017.

\bibitem{47}
Hao-Tian Wu, Ruoyan Jia, Jean-Luc Dugelay, and Junhui He.
\newblock Reversible image visual transformation for privacy and content
  protection.
\newblock {\em Multimedia Tools and Applications}, 80:30863--30877, 2021.

\bibitem{79}
Wenying Wen, Ziye Yuan, Shuren Qi, Yushu Zhang, and Yuming Fang.
\newblock {PPM-SEM}: A privacy-preserving mechanism for sharing electronic
  patient records and medical images in telemedicine.
\newblock {\em IEEE Transactions on Multimedia}, 26:5795--5806, 2024.

\bibitem{80}
Ping Ping, Pan Wei, Deyin Fu, Bobiao Guo, Olano~Teah Bloh, and Feng Xu.
\newblock {IMIH}: Imperceptible medical image hiding for secure healthcare.
\newblock {\em IEEE Transactions on Dependable and Secure Computing},
  21(5):4652--4667, 2024.

\bibitem{146}
Xun Lin, Yi~Yu, Zitong Yu, Ruohan Meng, Jiale Zhou, Ajian Liu, Yizhong Liu,
  Shuai Wang, Wenzhong Tang, Zhen Lei, and Alex Kot.
\newblock {HideMIA}: Hidden wavelet mining for privacy-enhancing medical image
  analysis.
\newblock In {\em Proceedings of the 32nd ACM International Conference on
  Multimedia}, page 8110–8119. Association for Computing Machinery, 2024.

\bibitem{83}
Michael Ryoo, Brandon Rothrock, Charles Fleming, and Hyun~Jong Yang.
\newblock Privacy-preserving human activity recognition from extreme low
  resolution.
\newblock In {\em Proceedings of the AAAI Conference on Artificial
  Intelligence}, volume~31, 2017.

\bibitem{84}
Lin Yuan, Wu~Chen, Xiao Pu, Yan Zhang, Hongbo Li, Yushu Zhang, Xinbo Gao, and
  Touradj Ebrahimi.
\newblock {PRO-Face C}: Privacy-preserving recognition of obfuscated face via
  feature compensation.
\newblock {\em IEEE Transactions on Information Forensics and Security},
  19:4930--4944, 2024.

\bibitem{86}
Mengmei Ye, Zhongze Tang, Huy Phan, Yi~Xie, Bo~Yuan, and Sheng Wei.
\newblock Visual privacy protection in mobile image recognition using
  protective perturbation.
\newblock In {\em Proceedings of the 13th ACM Multimedia Systems Conference},
  pages 164--176, 2022.

\bibitem{81}
Rakibul Hasan, Patrick Shaffer, David Crandall, Eman~T Apu~Kapadia, et~al.
\newblock Cartooning for enhanced privacy in lifelogging and streaming videos.
\newblock In {\em Proceedings of the IEEE Conference on Computer Vision and
  Pattern Recognition Workshops}, pages 29--38, 2017.

\bibitem{82}
Ad{\'a}m Erd{\'e}lyi, Tibor Bar{\'a}t, Patrick Valet, Thomas Winkler, and
  Bernhard Rinner.
\newblock Adaptive cartooning for privacy protection in camera networks.
\newblock In {\em 2014 11th IEEE international conference on advanced video and
  signal based surveillance (AVSS)}, pages 44--49, 2014.

\bibitem{85}
MaungMaung AprilPyone and Hitoshi Kiya.
\newblock Block-wise image transformation with secret key for adversarially
  robust defense.
\newblock {\em IEEE Transactions on Information Forensics and Security},
  16:2709--2723, 2021.

\bibitem{87}
Seyed-Mohsen Moosavi-Dezfooli, Alhussein Fawzi, Omar Fawzi, and Pascal
  Frossard.
\newblock Universal adversarial perturbations.
\newblock In {\em Proceedings of the IEEE Conference on Computer Vision and
  Pattern Recognition}, pages 1765--1773, 2017.

\bibitem{88}
Jingyi Cao, Bo~Liu, Yunqian Wen, Rong Xie, and Li~Song.
\newblock Personalized and invertible face de-identification by disentangled
  identity information manipulation.
\newblock In {\em Proceedings of the IEEE/CVF International Conference on
  Computer Vision}, pages 3334--3342, 2021.

\bibitem{89}
Yi-Lun Pan, Jun-Cheng Chen, and Ja-Ling Wu.
\newblock A multi-factor combinations enhanced reversible privacy protection
  system for facial images.
\newblock In {\em 2021 IEEE International Conference on Multimedia and Expo
  (ICME)}, pages 1--6, 2021.

\bibitem{90}
Hugo Proen{\c{c}}a.
\newblock {The UU-Net}: Reversible face de-identification for visual
  surveillance video footage.
\newblock {\em IEEE Transactions on Circuits and Systems for Video Technology},
  32(2):496--509, 2021.

\bibitem{91}
Dongze Li, Wei Wang, Kang Zhao, Jing Dong, and Tieniu Tan.
\newblock Riddle: Reversible and diversified de-identification with latent
  encryptor.
\newblock 2023.

\bibitem{92}
Yushu Zhang, Tao Wang, Ruoyu Zhao, Wenying Wen, and Youwen Zhu.
\newblock {RAPP}: Reversible privacy preservation for various face attributes.
\newblock {\em IEEE Transactions on Information Forensics and Security},
  18:3074--3087, 2023.

\bibitem{93}
Xiao He, Mingrui Zhu, Dongxin Chen, Nannan Wang, and Xinbo Gao.
\newblock Diff-privacy: Diffusion-based face privacy protection.
\newblock {\em IEEE Transactions on Circuits and Systems for Video Technology},
  2024.
\newblock doi: {10.1109/TCSVT.2024.3449290}.

\bibitem{104}
Yunqian Wen, Bo~Liu, Jingyi Cao, Rong Xie, Li~Song, and Zhu Li.
\newblock {IdentityMask}: Deep motion flow guided reversible face video
  de-identification.
\newblock {\em IEEE Transactions on Circuits and Systems for Video Technology},
  32(12):8353--8367, 2022.

\bibitem{136}
Jingyi Cao, Bo~Liu, Yunqian Wen, Rong Xie, and Li~Song.
\newblock Achieving privacy-preserving multi-view consistency with advanced
  3{D}-aware face de-identification.
\newblock In {\em Proceedings of the 5th ACM International Conference on
  Multimedia in Asia}, pages 1--7, 2023.

\bibitem{137}
Yunqian Wen, Bo~Liu, Jingyi Cao, Rong Xie, and Li~Song.
\newblock Divide and conquer: A two-step method for high quality face
  de-identification with model explainability.
\newblock In {\em Proceedings of the IEEE/CVF International Conference on
  Computer Vision}, pages 5148--5157, 2023.

\bibitem{94}
Tao Li and Lei Lin.
\newblock Anonymousnet: Natural face de-identification with measurable privacy.
\newblock In {\em Proceedings of the IEEE/CVF Conference on Computer Vision and
  Pattern Recognition Workshops}, pages 56--65, 2019.

\bibitem{95}
Seyed-Mohsen Moosavi-Dezfooli, Alhussein Fawzi, and Pascal Frossard.
\newblock Deepfool: A simple and accurate method to fool deep neural networks.
\newblock In {\em Proceedings of the IEEE Conference on Computer Vision and
  Pattern Recognition}, pages 2574--2582, 2016.

\bibitem{96}
Maxim Maximov, Ismail Elezi, and Laura Leal-Taix{\'e}.
\newblock Ciagan: Conditional identity anonymization generative adversarial
  networks.
\newblock In {\em Proceedings of the IEEE/CVF Conference on Computer Vision and
  Pattern Recognition}, pages 5447--5456, 2020.

\bibitem{97}
Yongxiang Li, Qianwen Lu, Qingchuan Tao, Xingbo Zhao, and Yanmei Yu.
\newblock {SF-GAN}: Face de-identification method without losing facial
  attribute information.
\newblock {\em IEEE Signal Processing Letters}, 28:1345--1349, 2021.

\bibitem{98}
Jia-Wei Chen, Li-Ju Chen, Chia-Mu Yu, and Chun-Shien Lu.
\newblock Perceptual indistinguishability-net (pi-net): Facial image
  obfuscation with manipulable semantics.
\newblock In {\em Proceedings of the IEEE/CVF Conference on Computer Vision and
  Pattern Recognition}, pages 6478--6487, 2021.

\bibitem{99}
Vahid Mirjalili, Sebastian Raschka, and Arun Ross.
\newblock {PrivacyNet}: Semi-adversarial networks for multi-attribute face
  privacy.
\newblock {\em IEEE Transactions on Image Processing}, 29:9400--9412, 2020.

\bibitem{100}
Hui-Po Wang, Tribhuvanesh Orekondy, and Mario Fritz.
\newblock Infoscrub: Towards attribute privacy by targeted obfuscation.
\newblock In {\em Proceedings of the IEEE/CVF Conference on Computer Vision and
  Pattern Recognition}, pages 3281--3289, 2021.

\bibitem{101}
Tao Wang, Yushu Zhang, Ruoyu Zhao, Wenying Wen, and Rushi Lan.
\newblock Identifiable face privacy protection via virtual identity
  transformation.
\newblock {\em IEEE Signal Processing Letters}, 30:773--777, 2023.

\bibitem{105}
Bach~Ngoc Kim, Jose Dolz, Pierre-Marc Jodoin, and Christian Desrosiers.
\newblock {Privacy-Net}: An adversarial approach for identity-obfuscated
  segmentation of medical images.
\newblock {\em IEEE Transactions on Medical Imaging}, 40(7):1737--1749, 2021.

\bibitem{102}
Yiyi Xie, Yuqian Zhou, Tao Wang, Wenying Wen, Shuang Yi, and Yushu Zhang.
\newblock Reversible gender privacy enhancement via adversarial perturbations.
\newblock {\em Neural Networks}, 172(C), 2024.

\bibitem{106}
Zhen Chen, Xiuli Chai, Zhihua Gan, Binjie Wang, and Yushu Zhang.
\newblock {RAE-VWP}: A reversible adversarial example-based privacy and
  copyright protection method of medical images for internet of medical things.
\newblock {\em IEEE Internet of Things Journal}, 11(11):20757--20768, 2024.

\bibitem{107}
Wentao Zhou, Yushu Zhang, Ruoyu Zhao, Shuang Yi, and Rushi Lan.
\newblock Adversarial thumbnail-preserving transformation for facial images
  based on {GAN}.
\newblock {\em IEEE Signal Processing Letters}, 30:1147--1151, 2023.

\bibitem{108}
Jiawei Zhang, Jinwei Wang, Hao Wang, and Xiangyang Luo.
\newblock {Self-Recoverable Adversarial Examples}: A new effective protection
  mechanism in social networks.
\newblock {\em IEEE Transactions on Circuits and Systems for Video Technology},
  33(2):562--574, 2022.

\bibitem{109}
Xiao Yang, Yinpeng Dong, Tianyu Pang, Hang Su, Jun Zhu, Yuefeng Chen, and Hui
  Xue.
\newblock Towards face encryption by generating adversarial identity masks.
\newblock In {\em Proceedings of the IEEE/CVF International Conference on
  Computer Vision}, pages 3897--3907, 2021.

\bibitem{110}
Yaoyao Zhong and Weihong Deng.
\newblock Opom: Customized invisible cloak towards face privacy protection.
\newblock {\em IEEE Transactions on Pattern Analysis and Machine Intelligence},
  45(3):3590--3603, 2022.

\bibitem{125}
Xuannan Liu, Yaoyao Zhong, Weihong Deng, Hongzhi Shi, Xingchen Cui, Yunfeng
  Yin, and Dongchao Wen.
\newblock Enhancing generalization of invisible facial privacy cloak via
  gradient accumulation.
\newblock In {\em ICASSP 2024-2024 IEEE International Conference on Acoustics,
  Speech and Signal Processing}, pages 5290--5294, 2024.

\bibitem{114}
Yikun Xu, Pengwen Dai, Zekun Li, Hongjun Wang, and Xiaochun Cao.
\newblock The best protection is attack: Fooling scene text recognition with
  minimal pixels.
\newblock {\em IEEE Transactions on Information Forensics and Security},
  18:1580--1595, 2023.

\bibitem{115}
Baoyu Liang, Chao Tong, Chao Lang, Qinglong Wang, Joel JP~C Rodrigues, and
  Sergei Kozlov.
\newblock Protecting image privacy through adversarial perturbation.
\newblock {\em Multimedia Tools and Applications}, 81(24):34759--34774, 2022.

\bibitem{116}
Efstathios Chatzikyriakidis, Christos Papaioannidis, and Ioannis Pitas.
\newblock Adversarial face de-identification.
\newblock In {\em 2019 IEEE International conference on image processing
  (ICIP)}, pages 684--688, 2019.

\bibitem{117}
Valeriia Cherepanova, Micah Goldblum, Harrison Foley, Shiyuan Duan, John
  Dickerson, Gavin Taylor, and Tom Goldstein.
\newblock Lowkey: Leveraging adversarial attacks to protect social media users
  from facial recognition.
\newblock In {\em International Conference on Learning Representations}, 2021.
\newblock doi: {10.48550/arXiv.2101.07922}.

\bibitem{118}
Saheb Chhabra, Richa Singh, Mayank Vatsa, and Gaurav Gupta.
\newblock Anonymizing k-facial attributes via adversarial perturbations.
\newblock page 656–662, 2018.

\bibitem{119}
Oran Gafni, Lior Wolf, and Yaniv Taigman.
\newblock Live face de-identification in video.
\newblock In {\em Proceedings of the IEEE/CVF International Conference on
  Computer Vision}, pages 9378--9387, 2019.

\bibitem{124}
Chau~Yi Li, Ali~Shahin Shamsabadi, Ricardo Sanchez-Matilla, Riccardo Mazzon,
  and Andrea Cavallaro.
\newblock Scene privacy protection.
\newblock In {\em ICASSP 2019-2019 IEEE International Conference on Acoustics,
  Speech and Signal Processing}, pages 2502--2506, 2019.

\bibitem{126}
Wen Sun, Jian Jin, and Weisi Lin.
\newblock Minimum noticeable difference-based adversarial privacy preserving
  image generation.
\newblock {\em IEEE Transactions on Circuits and Systems for Video Technology},
  33(3):1069--1081, 2022.

\bibitem{127}
Xiaojun Jia, Xingxing Wei, Xiaochun Cao, and Xiaoguang Han.
\newblock Adv-watermark: A novel watermark perturbation for adversarial
  examples.
\newblock In {\em Proceedings of the 28th ACM International Conference on
  Multimedia}, pages 1579--1587, 2020.

\bibitem{111}
Zuomin Qu, Zuping Xi, Wei Lu, Xiangyang Luo, Qian Wang, and Bin Li.
\newblock {DF-RAP}: A robust adversarial perturbation for defending against
  deepfakes in real-world social network scenarios.
\newblock {\em IEEE Transactions on Information Forensics and Security},
  19:3943--3957, 2024.

\bibitem{113}
Wanlun Ma, Derui Wang, Chao Chen, Sheng Wen, Gaolei Fei, and Yang Xiang.
\newblock {LocGuard}: A location privacy defender for image sharing.
\newblock {\em IEEE Transactions on Dependable and Secure Computing},
  21(6):5526--5537, 2024.

\bibitem{112}
Long Tang, Dengpan Ye, Yunna Lv, Chuanxi Chen, and Yunming Zhang.
\newblock {Once and for All}: Universal transferable adversarial perturbation
  against deep hashing-based facial image retrieval.
\newblock In {\em Proceedings of the AAAI Conference on Artificial
  Intelligence}, volume~38, pages 5136--5144, 2024.

\bibitem{123}
Linxi Jiang, Xingjun Ma, Shaoxiang Chen, James Bailey, and Yu-Gang Jiang.
\newblock Black-box adversarial attacks on video recognition models.
\newblock In {\em Proceedings of the 27th ACM International Conference on
  Multimedia}, pages 864--872, 2019.

\bibitem{120}
Guha Balakrishnan, Fredo Durand, and John Guttag.
\newblock Detecting pulse from head motions in video.
\newblock In {\em Proceedings of the IEEE Conference on Computer Vision and
  Pattern Recognition}, pages 3430--3437, 2013.

\bibitem{121}
Mingliang Chen, Xin Liao, and Min Wu.
\newblock {PulseEdit}: Editing physiological signals in facial videos for
  privacy protection.
\newblock {\em IEEE Transactions on Information Forensics and Security},
  17:457--471, 2022.

\bibitem{122}
Zhaodong Sun and Xiaobai Li.
\newblock Privacy-phys: Facial video-based physiological modification for
  privacy protection.
\newblock {\em IEEE Signal Processing Letters}, 29:1507--1511, 2022.

\bibitem{128}
Yueming Lyu, Yue Jiang, Ziwen He, Bo~Peng, Yunfan Liu, and Jing Dong.
\newblock {3D-Aware} adversarial makeup generation for facial privacy
  protection.
\newblock {\em IEEE Transactions on Pattern Analysis and Machine Intelligence},
  45(11):13438--13453, 2023.

\bibitem{130}
Fahad Shamshad, Muzammal Naseer, and Karthik Nandakumar.
\newblock Clip2protect: Protecting facial privacy using text-guided makeup via
  adversarial latent search.
\newblock In {\em Proceedings of the IEEE/CVF Conference on Computer Vision and
  Pattern Recognition}, pages 20595--20605, 2023.

\bibitem{138}
Hanyu Xue, Bo~Liu, Xin Yuan, Ming Ding, and Tianqing Zhu.
\newblock Face image de-identification by feature space adversarial
  perturbation.
\newblock {\em Concurrency and Computation: Practice and Experience},
  35(5):e7554, 2023.

\bibitem{134}
Ruoyu Zhao, Yushu Zhang, Tao Wang, Wenying Wen, Yong Xiang, and Xiaochun Cao.
\newblock Visual content privacy protection: A survey.
\newblock {\em arXiv preprint arXiv:2303.16552}, 2023.

\end{thebibliography}

\end{document}